\documentclass[a4paper,aps,prd,10pt,preprintnumbers,showpacs,twocolumn,superscriptaddress,nofootinbib,amsmath,amssymb]{revtex4-1}
\usepackage{graphicx}
\usepackage{cmap}
\usepackage[utf8]{inputenc}
\usepackage[T1]{fontenc}
\usepackage{xcolor}
\usepackage{hyperref}

\def\re#1{Re(#1)}
\def\im#1{Im(#1)}

\begin{document}
\title{Grey-body factors and absorption cross-sections of scalar and Dirac fields in the vicinity of dilaton-de Sitter black hole}

\author{Bekir Can Lütfüoğlu}
\email{bekir.lutfuoglu@uhk.cz}
\affiliation{Department of Physics, Faculty of Science, University of Hradec Králové, Rokitanského 62/26, 500 03 Hradec Králové, Czech Republic}

\begin{abstract}
We investigate the propagation of a massive scalar field and a massless Dirac field in the geometry of a dilaton--de Sitter black hole. Starting from the covariant perturbation equations, we present the corresponding effective potentials and analyze their dependence on the dilaton charge, field mass, and cosmological constant. Using the WKB approximation, we compute the grey-body factors and study the associated absorption cross-sections. The results show that increasing the field mass or dilaton charge raises the effective potential barrier, leading to a suppression of transmission at low frequencies, while a larger cosmological constant lowers the barrier and enhances transmission. The partial absorption cross-sections for different multipole numbers display the expected oscillatory structure, with the lowest multipoles dominating at small frequencies. After summation over multipoles, the oscillations average out and the total cross-section interpolates between strong suppression in the infrared regime and the geometric capture limit at high frequencies. These findings provide a systematic description of scattering and absorption properties of dilaton--de Sitter black holes for both scalar and fermionic perturbations.
\end{abstract}

\maketitle
\section{Introduction}

Black-hole grey-body factors and absorption cross-sections provide essential links between theoretical models and observable signatures of black-hole radiation. While quasinormal modes \cite{Konoplya:2011qq,Kokkotas:1999bd,Nollert:1999ji,Bolokhov:2025uxz} encode the response of a black hole to external perturbations, grey-body factors determine the modification of Hawking radiation spectra \cite{Hawking:1975vcx} due to the potential barriers surrounding the horizon \cite{Page:1976df,Page:1976ki,Kanti:2004nr}. These quantities directly influence the energy emission rates and spectral distributions of evaporating black holes, thereby playing a central role in connecting black-hole physics with astrophysical and cosmological observations. In particular, for spacetimes that deviate from pure Schwarzschild or Kerr geometry---such as black holes with non-trivial matter fields or a cosmological constant---grey-body factors offer a powerful diagnostic tool for distinguishing between competing models. Therefore, a great number of publications are dedicated to calculations of grey-body factors \cite{Konoplya:2025uta,Duffy:2005ns,Kanti:2002ge,Klebanov:1997cx,Konoplya:2023ppx,Pappas:2016ovo,Harris:2003eg,Bolokhov:2024voa,Konoplya:2020cbv,Cvetic:1997uw,Bonanno:2025dry,Kanti:2017ubd,Gubser:1996zp,Konoplya:2023moy,Konoplya:2020jgt,Dubinsky:2024nzo,Fernando:2016ksb,Malik:2024wvs,Stashko:2024wuq,Konoplya:2021ube,Lutfuoglu:2025ljm,Konoplya:2023ahd,Harris:2005jx,Pappas:2017kam,Cvetic:1997xv,Bolokhov:2023ozp,Kokkotas:2010zd,Kanti:2014vsa} Notice that while in Hawking radiation, the temperature is usually the dominant factor determining the intensity of evaporation, in some cases the grey-body factors play the crucial role \cite{Konoplya:2019ppy,Konoplya:2019xmn}.

Among the various extensions of general relativity, black holes arising in dilaton gravity attract special attention \cite{Konoplya:2001ji,Konoplya:2022zav,Fernando:2004ay,Fernando:2016ksb}. The coupling of the dilaton to gauge fields modifies both the near-horizon and asymptotic structure of the spacetime, leading to distinctive dynamical and thermodynamic features. When combined with a positive cosmological constant, the resulting dilaton--de Sitter black holes represent a natural framework for exploring the interplay between string-inspired scalar fields and the accelerating Universe. Analyzing grey-body factors in this background is thus crucial not only for understanding fundamental aspects of black-hole scattering, but also for assessing their potential observational signatures in the electromagnetic and gravitational-wave spectra \cite{Oshita:2023cjz,Konoplya:2024lir}.

The inclusion of different field spins is likewise motivated. Massive scalar perturbations probe long-lived configurations and quasi-resonant states, which can strongly affect the late-time signal, while Dirac fields provide insight into fermionic degrees of freedom and their role in black-hole evaporation. Importantly, the spectral features of these two cases complement each other, offering a broader picture of how matter interacts with dilaton--de Sitter black holes.

Finally, special attention must be paid to the regime of large field mass, when the dimensionless parameter
\[
\frac{\mu M}{M_{\rm Pl}} \gg 1,
\]
becomes much larger than unity. In this limit, the Compton wavelength of the particle is much smaller than the horizon scale, and the absorption cross-section approaches the classical capture cross-section determined by null geodesics. Exploring this transition from the quantum to the semiclassical regime provides a deeper understanding of how black-hole scattering phenomena interpolate between microscopic and macroscopic scales, and clarifies the limits of validity of quantum-field theoretic approaches in curved spacetime.

One of the most distinctive signatures of black-hole spacetimes is their spectrum of quasinormal modes. These damped oscillations, often referred to as the ``proper'' frequencies of a black hole, represent the response of the geometry to perturbations and form the dominant part of the gravitational-wave ringdown signal that can be detected by modern interferometers such as LIGO and Virgo~\cite{LIGOScientific:2016aoc,LIGOScientific:2017vwq,LIGOScientific:2020zkf,Babak:2017tow}.

In asymptotically flat backgrounds, quasinormal ringing characterizes the intermediate stage of evolution and is eventually replaced at very late times by power-law tails. By contrast, for asymptotically de Sitter geometries, quasinormal oscillations determine not only the transient dynamics but also the asymptotic late-time behavior as $t\to\infty$~\cite{Dyatlov:2011jd,Dyatlov:2010hq,Konoplya:2024ptj}. This fundamental distinction makes de Sitter black holes particularly important in cosmological settings.

Over the years, a wide variety of dilaton-inspired black-hole models have been explored in this context. The quasinormal spectrum has been analyzed for spacetimes with minimally and non-minimally coupled dilaton fields, for different background asymptotics, and in extensions where additional corrections are included~\cite{Konoplya:2001ji,Fernando:2003wc,Chen:2005rm,Lopez-Ortega:2005obq,Lopez-Ortega:2009jpx,Zinhailo:2019rwd,Ferrari:2000ep,Carson:2020ter,Malybayev:2021lfq,Paul:2023eep,Blazquez-Salcedo:2020caw,Pierini:2022eim}. More recent investigations also considered dilatonic black holes embedded in higher-curvature frameworks, highlighting how string-inspired corrections and quantum modifications imprint themselves on the oscillation spectrum. Quasinormal modes of scalar and Dirac fields in the background of dilaton-de Sitter black hole were studied in detail in \cite{Dubinsky:2024hmn}, while some initial study of a scalar field perturbations was suggested in \cite{Fernando:2016ftj}. 

Although quasinormal modes and grey-body factors are defined through different boundary conditions, it has been shown in \cite{Konoplya:2024lir} that a correspondence exists between them. This relation is exact in the eikonal regime and only approximate outside of it. Furthermore, in certain cases \cite{Konoplya:2025ixm,Konoplya:2025hgp}, the correspondence breaks down even at a qualitative level. In this work, we take advantage of our analysis of dilaton–de Sitter black holes to compute their grey-body factors and, at the same time, assess the validity and accuracy of this correspondence. In this context, we use and confirm the data of \cite{Dubinsky:2024hmn} for quasinormal modes which were computed with sufficient accuracy.

The paper is organized as follows. In Sec.~\ref{sec:wavelike} we briefly review the dilaton--de Sitter black-hole solution and derive the master equations governing massive scalar and massless Dirac perturbations. The corresponding effective potentials are analyzed and their dependence on the model parameters is discussed. In Sec.~\ref{sec:GBF} we apply the WKB method to compute the grey-body factors and examine their behavior under variations of the dilaton charge, field mass, and cosmological constant. Sect.~\ref{sec:correspondence} is devoted to the correspondence between quasinormal modes and grey-body factors, with emphasis on its validity and limitations for the present background. Section~\ref{sec:absorption} is devoted to the study of absorption cross-sections, both partial and total, obtained by summing over multipole contributions. Finally, in Sec.~\ref{sec:conclusions} we summarize the main results and outline possible directions for further work.

\section{The dilaton-de Sitter black hole metric and wave equations}\label{sec:wavelike}

One of the earliest charged black-hole solutions in dilaton gravity was obtained by Gibbons and Maeda~\cite{Gibbons:1987ps}, and later rediscovered independently by Garfinkle, Horowitz, and Strominger~\cite{Garfinkle:1990qj}. This solution, which arises in the low-energy limit of string theory, is nowadays referred to as the Gibbons--Maeda--Garfinkle--Horowitz--Strominger (GMGHS) black hole. In the absence of a dilaton potential, asymptotically de Sitter solutions cannot be realized.

For a nontrivial dilaton potential, the action takes the form
\begin{equation}
S=\int d^4x \,\sqrt{-g}\,\Bigl[R - 2 (\nabla \Phi)^2 - V(\Phi) - e^{-2\Phi}F_{\mu\nu}F^{\mu\nu}\Bigr],
\end{equation}
where $R$ is the Ricci scalar, $F_{\mu\nu}$ the Maxwell field strength, and $\Phi$ the dilaton. The potential considered in~\cite{Gao:2004tu} is
\begin{equation}\label{potential}
V(\Phi)=\frac{\Lambda}{3}\left(e^{2(\Phi-\Phi_0)}+e^{-2(\Phi-\Phi_0)}\right)+\frac{4\Lambda}{3},
\end{equation}
with $\Lambda$ denoting the cosmological constant and $\Phi_0$ the asymptotic value of the dilaton.

The corresponding exact black-hole solution is
\begin{equation}\label{metric}
ds^2=-f(r)\,dt^2+\frac{dr^2}{f(r)}+R^2(r)\bigl(d\theta^2+\sin^2\theta\,d\phi^2\bigr),
\end{equation}
where the metric functions are
\begin{equation}
f(r)=1-\frac{2M}{r}-\frac{\Lambda r}{3}(r-2Q), \qquad R^2(r)=r^2-2Q r.
\end{equation}
Here $M$ is the black-hole mass and $Q$ the dilaton charge. When $\Lambda=0$, the line element reduces to the GMGHS black hole~\cite{Gibbons:1987ps,Garfinkle:1990qj}; when $Q=0$, one recovers the Schwarzschild--de Sitter geometry. In what follows we set $M=1$.  The accompanying matter fields are determined as
\begin{equation}
e^{2\Phi}=e^{2\Phi_0}\left(1-\frac{2Q}{r}\right), \quad
Q=\frac{q^2e^{2\Phi_0}}{2M}, \quad
F_{01}=\frac{qe^{2\Phi_0}}{r^2},
\end{equation}
where $q$ is the electric charge of the black hole.

\subsection*{Wave equations}

The dynamics of probe fields in this background are governed by the covariant Klein–Gordon and Dirac equations. For a massive scalar $\phi$ and a massless Dirac spinor $\Upsilon$, one has
\begin{subequations}\label{coveqs}
\begin{align}
\frac{1}{\sqrt{-g}}\partial_\mu\!\left(\sqrt{-g}\,g^{\mu\nu}\partial_\nu \phi\right)&= \mu^2\phi, \label{KGg}\\
\gamma^{\alpha}\!\left(\partial_\alpha-\Gamma_\alpha\right)\Upsilon&=0, \label{covdirac}
\end{align}
\end{subequations}
where $\mu$ is the scalar-field mass, $\gamma^\alpha$ are curved-space gamma matrices, and $\Gamma_\alpha$ the spin connections in the tetrad formalism.

After separation of variables and introduction of a suitable radial function $\Psi$, both equations reduce to a Schrödinger-like form~\cite{Kokkotas:1999bd,Berti:2009kk,Konoplya:2011qq}:
\begin{equation}\label{wave-equation}
\frac{d^2\Psi}{dr_*^2}+\bigl(\omega^2-V(r)\bigr)\Psi=0,
\end{equation}
with the tortoise coordinate defined as
\begin{equation}\label{tortoise}
dr_*\equiv\frac{dr}{f(r)}.
\end{equation}

For the scalar ($s=0$) case, the effective potential reads (see, for instance \cite{Abdalla:2006qj,Konoplya:2006rv})
\begin{equation}\label{potentialScalar}
V(r)=f(r)\left(\frac{\ell(\ell+1)}{R^2(r)}+\mu^2\right)+\frac{1}{R(r)}\frac{d^2R(r)}{dr_*^2},
\end{equation}
with $\ell=0,1,2,\dots$ the multipole number.

For the Dirac field, following \cite{Brill:1957fx}, one obtains two supersymmetric partner potentials
\begin{equation}
V_\pm(r)=W^2\pm\frac{dW}{dr_*}, \qquad
W=\left(\ell+\tfrac{1}{2}\right)\frac{\sqrt{f(r)}}{R(r)},
\end{equation}
where $\ell=1/2,3/2,\dots$. The associated wave functions $\Psi_\pm$ are related by a Darboux transformation:
\begin{equation}\label{psi}
\Psi_+\propto\Bigl(W+\frac{d}{dr_*}\Bigr)\Psi_-.
\end{equation}
Since $V_+$ and $V_-$ are isospectral, in practice it suffices to study quasinormal modes using only $V_+(r)$, where the WKB method is typically more accurate.

The plots of effective potentials are given in Figs. 3-8 of \cite{Dubinsky:2024hmn}. Therefore, we will not repeat them here.  From the analysis of the figures of effective potentials shown in Figs. 3-8 of \cite{Dubinsky:2024hmn} it becomes clear that increasing either the dilaton charge $Q$ or the field mass $\mu$ results in a pronounced growth of the maximum of the effective potential. The role of $\mu$ here is qualitatively different from what is observed in asymptotically flat geometries, where the potential asymptotes to a constant value at spatial infinity and, for sufficiently large $\mu$, no potential barrier forms. In contrast, within the dilaton--de Sitter background, the presence of a nonzero $\mu$ always generates a distinct barrier. For small $\mu$, an additional feature appears: a shallow negative well develops at a radial position close to the cosmological horizon. On the other hand, increasing the cosmological constant $\Lambda$ tends to lower the overall peak of the potential \cite{Dubinsky:2024hmn}.

\begin{figure*}
\resizebox{\linewidth}{!}{\includegraphics{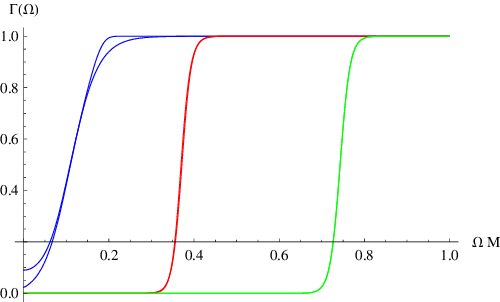}~~\includegraphics{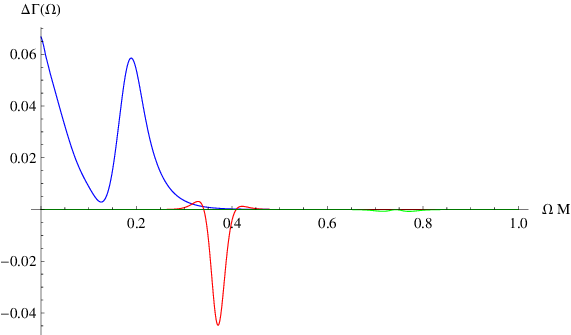}}
\caption{Left: Grey-body factors of a scalar field obtained by the 6th order WKB method and via the correspondence with the QNMs for $\Lambda=0.01$, $Q=0.01$, $\ell=0$:  $\mu=0$ (blue),
$\mu=0.5$ (red), and $\mu=1$ (green). Right: Difference between GBFs obtained by the  WKB data via the correspondence for the same values of the parameters. }\label{fig:L0-GBF}
\end{figure*}

\begin{figure*}
\resizebox{\linewidth}{!}{\includegraphics{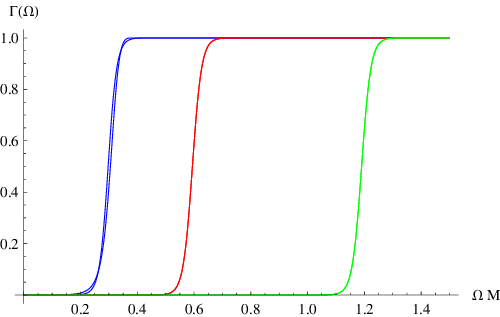}~~\includegraphics{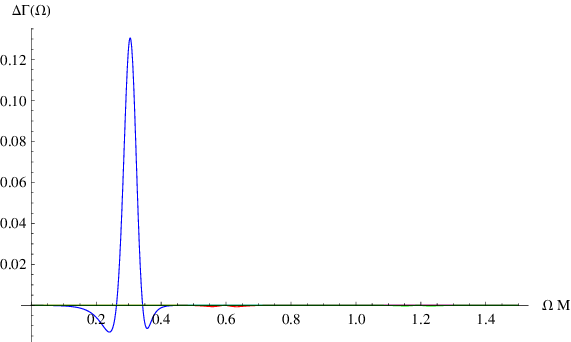}}
\caption{Left: Grey-body factors of a scalar field obtained by the 6th order WKB method and via the correspondence with the QNMs for $\Lambda=0.03$, $Q=0.01$, $\ell=0$:  $\mu=0.5$ (blue),
$\mu=1$ (red), and $\mu=2$ (green). Right: Difference between GBFs obtained by the  WKB data via the correspondence for the same values of the parameters. The difference $\Delta \Gamma$ for $\mu=1$ and $\mu=2$ is of the order $10^{-4}$ or smaller and cannot be seen on the plot.}\label{fig:L0-GBF2}
\end{figure*}

\begin{figure*}
\resizebox{\linewidth}{!}{\includegraphics{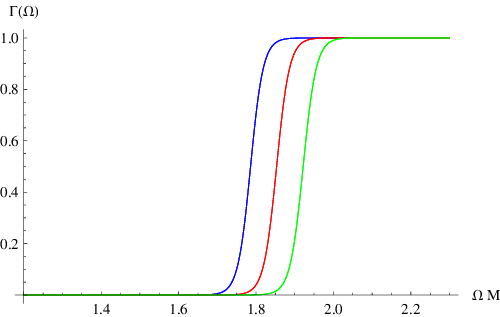}~~\includegraphics{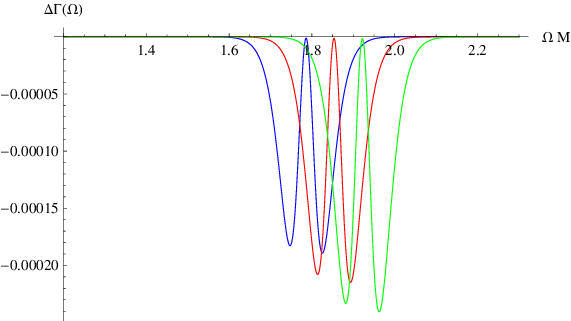}}
\caption{Left: Grey-body factors of a scalar field obtained by the 6th order WKB method and via the correspondence with the QNMs for $\Lambda=0.03$, $\mu=3$, $\ell=0$:  $Q=0.01$ (blue),
$Q=0.3$ (red), and $Q=0.6$ (green). Right: Difference between GBFs obtained by the  WKB data via the correspondence for the same values of the parameters.}\label{fig:L0-GBFQvar}
\end{figure*}

\begin{figure*}
\resizebox{\linewidth}{!}{\includegraphics{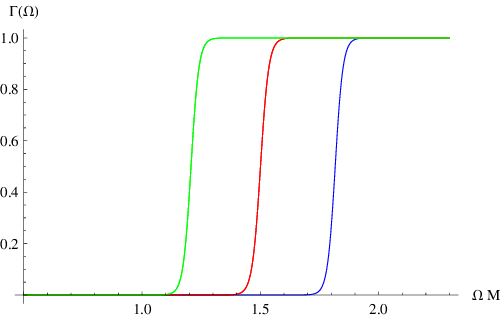}~~\includegraphics{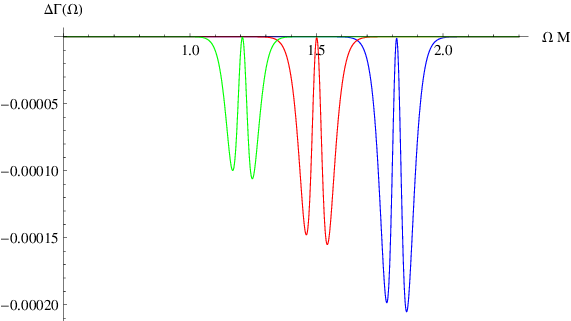}}
\caption{Left: Grey-body factors of a scalar field obtained by the 6th order WKB method and via the correspondence with the QNMs for $Q=0.1$, $\mu=3$, $\ell=1$: $\Lambda=0.03$ (blue), $\Lambda=0.05$ (red), and $\Lambda=0.07$ (green). Right: Difference between GBFs obtained by the  WKB data via the correspondence for the same values of the parameters.}\label{fig:L1-GBFLambdaVar}
\end{figure*}

\begin{figure*}
\resizebox{\linewidth}{!}{\includegraphics{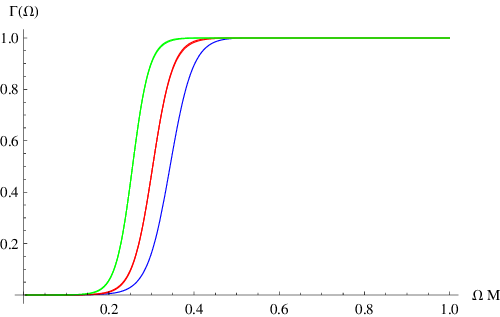}~~\includegraphics{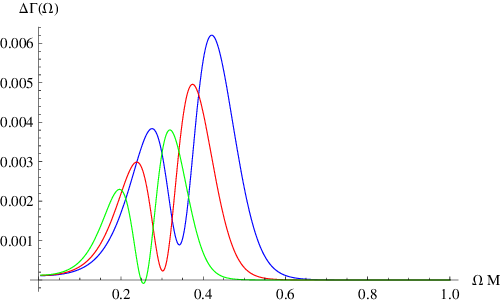}}
\caption{Left: Grey-body factors of a Dirac field obtained by the 6th order WKB method and via the correspondence with the QNMs for $Q=0.1$, $\mu=3$, $\ell=3/2$: $\Lambda=0.03$ (blue), $\Lambda=0.05$ (red), and $\Lambda=0.07$ (green). Right: Difference between GBFs obtained by the  WKB data via the correspondence for the same values of the parameters.}\label{fig:Dirac-GBFLambdaVar}
\end{figure*}


\begin{figure*}
\resizebox{\linewidth}{!}{\includegraphics{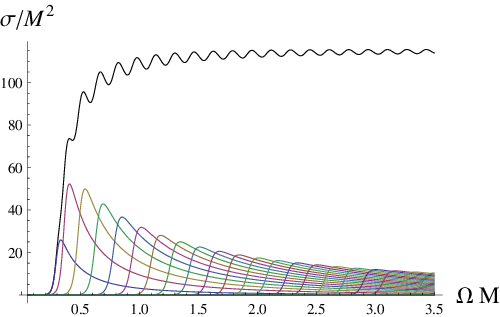}\includegraphics{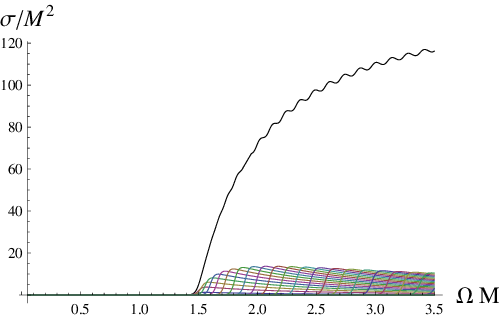}\includegraphics{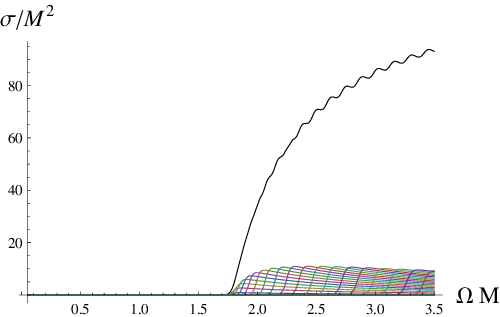}}
\caption{Absorption cross-section of a scalar field for the first fifty multipole numbers together with the total cross-section for $\Lambda=0.03$, $Q=0.01$, $\mu=0.5$ (left), $\mu=1.5$ (middle) and $\mu=3$ (right). Only a few tens of these multipoles contribute at smaller frequencies.} \label{fig:sigmaScalar}
\end{figure*}

\begin{figure*}
\resizebox{\linewidth}{!}{\includegraphics{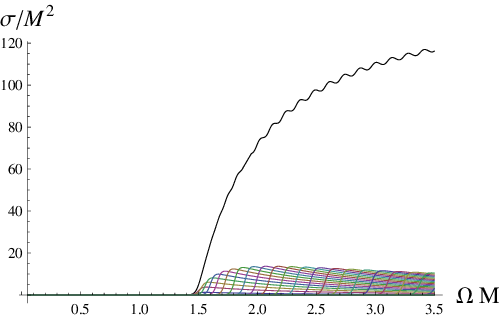}\includegraphics{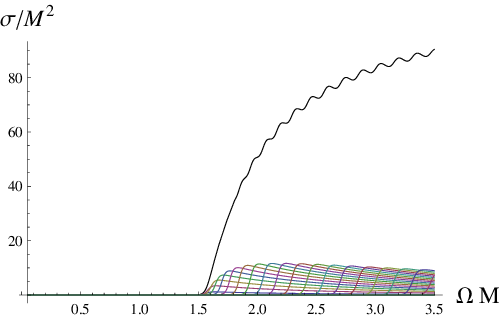}\includegraphics{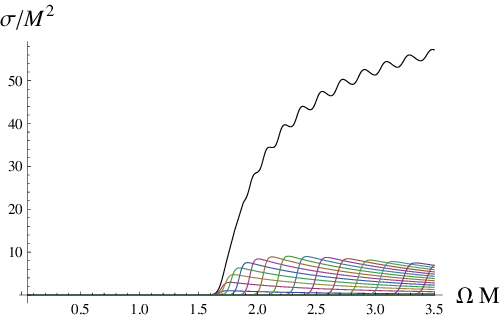}}
\caption{Absorption cross-section of a scalar field for the first fifty multipole numbers together with the total cross-section for $\Lambda=0.05$, $\mu=3$ and $Q=0.1$ (left), $Q=0.3$ (middle) and $Q=0.6$ (right). Only a few tens of these multipoles contribute at smaller frequencies.}\label{fig:sigmaScalarVarQ}
\end{figure*}

\begin{figure*}
\resizebox{\linewidth}{!}{\includegraphics{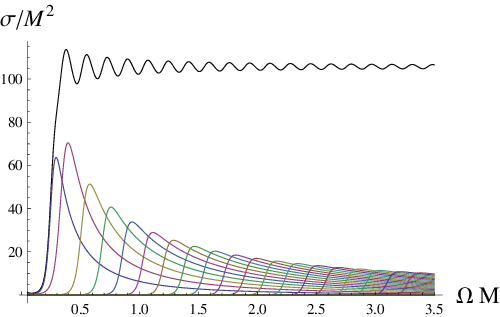}~~\includegraphics{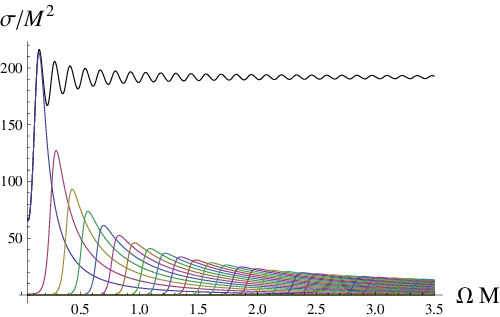}~~\includegraphics{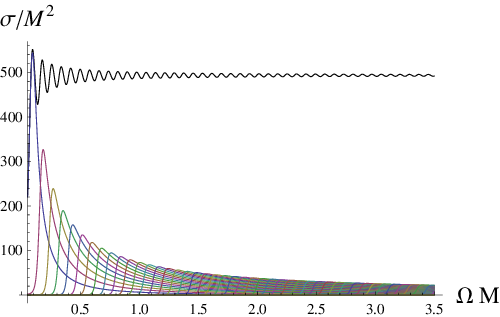}}
\caption{Absorption cross-sections of a Dirac field for the first fifty multipole numbers together with the total cross-section for  $Q=0.1$: $\Lambda=0.03$ (left), $\Lambda=0.07$ (middle), and $\Lambda=0.1$,  (right). Only a few tens of these multipoles contribute at smaller frequencies.}\label{fig:Dirac-SigmaLambdaVar}
\end{figure*}
\vspace{4mm}

\section{Grey-body factors}\label{sec:GBF}

The radiation emitted by black holes is not purely thermal but is filtered by the effective potential surrounding the horizon.
The transmission probability through this potential barrier is quantified by the so-called grey-body factors (GBFs), which play a crucial role in determining observable spectra of Hawking radiation.
Although asymptotically de Sitter spacetimes differ globally from asymptotically flat ones, the associated scattering problem in terms of the tortoise coordinate $r^{*}$ is effectively the same: the near-horizon region $r^{*}\to -\infty$ corresponds to the black-hole horizon in the asymptotically flat case, while in the de Sitter case it corresponds either to the black-hole or to the cosmological horizon.
Therefore, the standard boundary conditions and methods apply to both classes of geometries.

The scattering problem is governed by the Schrödinger-like wave equation \ref{wave-equation},
where $\Omega$ is the \textit{real} frequency of the radiation experiencing scattering and $V(r)$ is the effective potential depending on the type of field.

The physically motivated boundary conditions require purely ingoing waves at the event horizon and a superposition of ingoing and outgoing waves at spatial infinity (or at the cosmological horizon in the de Sitter case). Actual processes corresponding to scattering of radiation going from the event horizon towards the peak of the potential usually require opposite boundary conditions with the mixture of in- and out- waves near the horizon. However, due to the symmetry of $S$-matrix we can reformulate it as follows:
\begin{widetext}
\begin{equation}
\Psi(r^*) \sim
\begin{cases}
A_{\rm in}\,e^{-i\Omega r^*}+A_{\rm out}\,e^{+i\Omega r^*}, & r^*\to +\infty, \\[6pt]
B_{\rm in}\,e^{-i\Omega r^*}, & r^*\to -\infty \ \text{(event horizon)} .
\end{cases}
\end{equation}
\end{widetext}
From this, the amplitude ratios are defined as
\begin{equation}
\mathcal R_\ell(\Omega)=\frac{A_{\rm out}}{A_{\rm in}},
\qquad
\mathcal T_\ell(\Omega)=\frac{B_{\rm in}}{A_{\rm in}} .
\end{equation}

The corresponding reflection and transmission coefficients (probabilities),
which are determined by flux ratios, take the form
\begin{equation}
R_\ell(\Omega)=|\mathcal R_\ell|^2=\frac{|A_{\rm out}|^2}{|A_{\rm in}|^2},
\quad
\Gamma_\ell(\Omega)=|\mathcal T_\ell|^2=\frac{|B_{\rm in}|^2}{|A_{\rm in}|^2}.
\end{equation}

For real effective potentials, conservation of flux implies
\begin{equation}
R_\ell(\Omega) + \Gamma_\ell(\Omega) = 1.
\end{equation}
Thus, the grey-body factor coincides with the transmission probability,
\begin{equation}
\Gamma_\ell(\Omega)=1-\frac{|A_{\rm out}|^2}{|A_{\rm in}|^2}.
\end{equation}
Since the effective potentials typically form a barrier with a single maximum, the Wentzel–Kramers–Brillouin (WKB) method provides a convenient way to approximate the transmission probability.
In the $N$-th order WKB expansion~\cite{Schutz:1985km,Iyer:1986np,Konoplya:2003ii,Konoplya:2019hlu}, the grey-body factor for a given multipole $\ell$ is expressed as
\begin{equation}\label{gbf-WKB}
\Gamma_\ell(\Omega) \approx \frac{1}{1 + e^{2\pi K}},
\end{equation}
where
\begin{equation}
K =\sum_{k=1}^N \Lambda_k(\{V_0^{(j)}\}, \Omega).
\end{equation}
Here $V_0$ is the maximum of the effective potential, $V_0^{(j)}$ denotes its $j$-th derivative with respect to $r^{*}$ at the peak, and $\Lambda_k$ are higher-order WKB correction terms.  For $N$-th WKB order the maximal value of $j=2 N$. The explicit form of the above WKB formula up to the 13th order can be found in  \cite{Schutz:1985km,Iyer:1986np,Konoplya:2003ii,Konoplya:2019hlu}. Notice that the WKB method provides sufficient accuracy in the majority of cases for the dominant quasinormal modes and grey-body factors (see, \cite{Bolokhov:2023ruj,Bolokhov:2023bwm,Skvortsova:2024wly,Skvortsova:2024eqi,Balart:2023odm,Malik:2023bxc,Malik:2025ava,Malik:2024nhy,Al-Badawi:2023lvx,Chen:2023akf,Bolokhov:2023dxq,Bolokhov:2024ixe,Skvortsova:2023zmj,Skvortsova:2024atk,Zinhailo:2019rwd,Zinhailo:2018ska,Churilova:2021tgn,Bronnikov:2021liv,Paul:2023eep} for recent examples).

Grey-body factors for various values of parameters are shown in Figs. \ref{fig:L0-GBF}-\ref{fig:Dirac-GBFLambdaVar}.
Our analysis shows that the behavior of the effective potential and the corresponding grey-body factors is strongly influenced by the parameters of the system. An increase in the cosmological constant $\Lambda$ lowers the height of the potential barrier, leading to a noticeable enhancement of the grey-body factors (see Figs. \ref{fig:L1-GBFLambdaVar} and \ref{fig:Dirac-GBFLambdaVar}). In contrast, increasing either the dilaton charge $Q$ or the field mass $\mu$ raises the peak of the potential and, as a result, suppresses the transmission probability, as shown in Figs. \ref{fig:L0-GBF}, \ref{fig:L0-GBF2}, and \ref{fig:L0-GBFQvar}. In both cases (of increasing either $Q$ or $\mu$), the low-frequency part of the spectrum is strongly suppressed, so that only waves with sufficiently high frequencies pass the barrier. For larger values of $\mu$, this suppression becomes especially pronounced, effectively shifting the emission spectrum toward higher frequencies.\\

\section{Correspondence between Quasinormal Modes and Grey-Body Factors}\label{sec:correspondence}

Konoplya and Zhidenko~\cite{Konoplya:2024lir} have recently demonstrated that, within the WKB framework, grey-body factors can be directly expressed in terms of the lowest quasinormal frequencies. In the eikonal regime, the transmission probability takes the universal form
\begin{eqnarray}
\Gamma_\ell(\Omega) &\simeq& \left[1+\exp\!\left(2\pi\,K(\Omega)\right)\right]^{-1},
\\\nonumber
K(\Omega) &\simeq& \frac{\Omega^{2}-\re{\omega_{0}}^{2}}
{4\,\re{\omega_{0}}\,|\im{\omega_{0}}|},
\end{eqnarray}
where $\omega_{0}$ is the fundamental quasinormal frequency. This relation becomes exact for large $\ell$, linking the peak-barrier scattering to the dominant ringdown mode.

The quasinormal frequencies $\omega$, in turn, obey different boundary conditions: they correspond to purely ingoing waves at the event horizon and purely outgoing waves at spatial infinity (or at the de Sitter horizon in the asymptotically de Sitter case). These are scattering-type boundary conditions, which are equivalent to the poles (divergences) of the reflection coefficient, i.e., to the case $A_{\rm in}=0$.

For moderate multipoles, corrections enter through higher-order WKB terms and, in particular, through the first overtone $\omega_{1}$, leading to improved approximations of the form
\begin{equation}
\Gamma_\ell(\Omega) \simeq
\left[1+\exp\!\Bigl(2\pi \bigl[K(\Omega)+\Delta_1+\Delta_2+\dots \bigr]\Bigr)\right]^{-1},
\end{equation}
with $\Delta_{i}$ expressed via $\omega_{0}$ and $\omega_{1}$. These refinements substantially enhance the accuracy at smaller $\ell$.

The correspondence illustrates that grey-body factors are largely governed by the lowest QNMs, which are comparatively stable against moderate modifications of the effective potential. The correspondence has been recently tested and applied in a few works \cite{Han:2025cal,Bolokhov:2024otn,Skvortsova:2024msa,Malik:2024cgb,Lutfuoglu:2025hjy,Lutfuoglu:2025blw,Malik:2025dxn,Dubinsky:2024vbn,Bolokhov:2025lnt,Lutfuoglu:2025ohb,Lutfuoglu:2025ldc,Dubinsky:2025nxv,Malik:2025erb}. Nonetheless, the relation ceases to hold in scenarios where the potential develops multiple peaks or departs from the WKB-good form, underscoring its range of applicability. For example, in several theories with higher-curvature corrections \cite{Konoplya:2017lhs,Konoplya:2017wot}, the correspondence between eikonal quasinormal modes and null geodesics breaks down, since the effective potential in the eikonal limit no longer takes the usual centrifugal form $\sim g_{tt}\,\ell(\ell+1)/r^{2}$. In such cases, the analogous correspondence between eikonal quasinormal modes and grey-body factors also fails.

An additional layer of complexity in the QNM--GBF correspondence arises when a non-vanishing cosmological constant is taken into account.
In asymptotically de Sitter spacetimes, the quasinormal spectrum does not form a single  family of modes, but rather separates into two distinct branches. One branch behaves perturbatively in $\Lambda$, smoothly approaching the Schwarzschild quasinormal spectrum in the limit $\Lambda \to 0$~\cite{Konoplya:2004uk,Zhidenko:2003wq,Moss:2001ga}.
The second branch, in contrast, is non-perturbative in nature and, for black holes whose horizon radius is much smaller than the de Sitter scale, asymptotes to the characteristic frequencies of pure de Sitter space~\cite{Lopez-Ortega:2007vlo,Lopez-Ortega:2012xvr}. For this de Sitter branch of modes the correspondence with null geodesics is not valid, as shown in \cite{Konoplya:2022gjp}.

It is also important to contrast this situation with asymptotically flat geometries. There, massive fields of different spins may exhibit arbitrarily long-lived oscillations—so-called quasi-resonances—whose lifetime diverges at specific values of the mass~\cite{Konoplya:2006gq,Ohashi:2004wr}.
Once a cosmological constant is introduced, however, this phenomenon disappears: no quasi-resonant states emerge, and the spectrum is instead governed by the two de Sitter-related branches described above.

We can thus conclude that the correspondence is valid for the first, perturbative branch, but fails — even approximately — for the non-perturbative one. When the multipole number $\ell$ is increased, the accuracy of the correspondence increases quickly.\\

\section{Absorption Cross-Section}\label{sec:absorption}

The absorption cross-section of a black hole provides a measure of its effective size as seen by incident waves. It quantifies the probability that incoming radiation of a given frequency $\Omega$ and multipole number $\ell$ will penetrate the effective potential barrier and be absorbed by the horizon, rather than scattered back to infinity. Once the grey-body factors $\Gamma_{\ell}(\Omega)$ are known, the total absorption cross-section can be obtained by summing over all partial waves \cite{Futterman:1988ni}:
\begin{equation}
\sigma(\Omega) \;=\; \frac{\pi}{\Omega^{2}} \sum_{\ell=0}^{\infty} (2\ell+1)\, \Gamma_{\ell}(\Omega).
\end{equation}
This expression resembles the optical theorem in quantum scattering, with the grey-body factors playing the role of transmission probabilities for each angular momentum mode.

Physically, $\sigma(\Omega)$ can be interpreted as the ``effective area'' of the black hole that interacts with an incident wave of frequency $\Omega$. In the low-frequency limit, $\sigma(\Omega)$ approaches the geometric area of the event horizon, while in the high-frequency (eikonal) regime it converges to the classical capture cross-section determined by null geodesics. At intermediate frequencies, however, the cross-section exhibits rich structure due to the interference of partial waves, giving rise to oscillatory patterns often referred to as ``glory scattering.'' Thus, the absorption cross-section provides a bridge between quantum-wave effects and the classical description of particle capture, offering a powerful diagnostic of black-hole geometry and possible deviations from general relativity.

In Figs.~6--8 the absorption cross-sections are decomposed into partial contributions of the first ten multipole numbers $\ell$ and then summed to obtain the total spectrum. Each individual partial cross-section displays the expected oscillatory structure, which arises from interference between waves scattered by the effective potential barrier. At low frequencies the absorption is dominated by the lowest multipoles, while higher-$\ell$ modes contribute only at larger frequencies, as expected from the fact that short-wavelength modes are able to probe the barrier more efficiently. 

In addition, the field mass $\mu$ has a distinctive effect: larger $\mu$ raises the effective potential barrier, leading to a stronger suppression of the absorption cross-section at low frequencies and effectively shifting the onset of significant absorption toward higher frequencies (see Fig. 6). This trend is consistent with the fact that massive fields experience an additional potential barrier at large distances, which reduces their transmission probability. The influence of the dilaton charge $Q$ is also pronounced: increasing $Q$ enhances the effective potential, which results in a lower overall absorption cross-section. This suppression is most significant at low frequencies, while at higher frequencies the effect of $Q$ becomes weaker, as the radiation is able to overcome the barrier more efficiently. A clear influence of the cosmological constant $\Lambda$ can be observed in these plots. Increasing $\Lambda$ generally lowers the effective potential barrier, leading to an overall growth of the gray-body factors and, consequently, to an improvement in the absorption cross-section over all multipole numbers (see Fig. 8). The Dirac field (Fig.~8) shows qualitatively similar behavior, but with a richer oscillatory pattern in the partial contributions, which partially survives even after summation.

Although the canonical formula for the absorption cross section was first obtained for asymptotically flat black holes, it can be consistently generalized to the de Sitter case. This is possible because, when written in terms of the tortoise coordinate $r_{*}$, the perturbation equation retains its Schrödinger-like structure, with $r_{*}$ covering the entire real line from $-\infty$ to $+\infty$. The essential distinction is geometric rather than formal: the region accessible to wave propagation is now limited by two horizons—the event and the cosmological ones—so that the effective scattering domain is finite.

When summed over several multipoles, the oscillations from different $\ell$ channels tend to average out, producing a smoother absorption profile. The total cross-section then interpolates between two limiting regimes: at low frequencies it is strongly suppressed, reflecting the dominance of the barrier, while at high frequencies it approaches the geometric capture cross-section. The position of the intermediate maximum is governed by the effective barrier structure and depends sensitively on the values of $\Lambda$, $Q$, and $\mu$, providing a useful diagnostic of the underlying black-hole spacetime.

\section{Conclusions}\label{sec:conclusions}

In this paper we have studied the propagation of massive scalar and massless Dirac fields in the background of dilaton--de Sitter black holes. Starting from the covariant perturbation equations, we presented the effective wave-like equations and potentials, and analyzed their qualitative dependence on the field mass $\mu$, the dilaton charge $Q$, and the cosmological constant $\Lambda$.

We calculated the grey-body factors using the WKB approach and showed that they are strongly suppressed at low frequencies when either $\mu$ or $Q$ is increased, while a larger cosmological constant reduces the barrier and tends to enhance transmission. This behavior translates directly into the absorption cross-sections. Decomposing the absorption into partial contributions for the first ten multipole numbers, we found that low-frequency absorption is dominated by the lowest multipoles, whereas higher $\ell$–modes contribute only at larger frequencies. The mass of the scalar field shifts the onset of significant absorption toward higher frequencies, the dilaton charge suppresses absorption across all regimes, and the cosmological constant lowers the overall cross-section. For the Dirac field, the qualitative behavior is similar, though with a richer oscillatory structure in the partial cross-sections.

When summed over multipoles, the oscillations largely average out, yielding smooth total absorption spectra that interpolate between two limiting regimes: strong suppression at low frequencies and the geometric capture cross-section at high frequencies. These results provide a consistent picture of how the parameters $\mu$, $Q$, and $\Lambda$ shape the scattering and absorption properties of dilaton--de Sitter black holes.

Our work could be extended to the coupled gravitational and electromagnetic perturbations. However, the derivation of the wave-like equations is a complex problem in this case.

\begin{acknowledgments}
B. C. L. is grateful to the Excellence project FoS UHK 2205/2025-2026 for the financial support.
\end{acknowledgments}

\section*{Conflict of Interest}
The authors declare no competing interests.

\section*{Data Availability Statement}
All data generated or analyzed during this study are included in this published article.

\bibliography{bibliography}

\begin{thebibliography}{117}%
\makeatletter
\providecommand \@ifxundefined [1]{%
 \@ifx{#1\undefined}
}%
\providecommand \@ifnum [1]{%
 \ifnum #1\expandafter \@firstoftwo
 \else \expandafter \@secondoftwo
 \fi
}%
\providecommand \@ifx [1]{%
 \ifx #1\expandafter \@firstoftwo
 \else \expandafter \@secondoftwo
 \fi
}%
\providecommand \natexlab [1]{#1}%
\providecommand \enquote  [1]{``#1''}%
\providecommand \bibnamefont  [1]{#1}%
\providecommand \bibfnamefont [1]{#1}%
\providecommand \citenamefont [1]{#1}%
\providecommand \href@noop [0]{\@secondoftwo}%
\providecommand \href [0]{\begingroup \@sanitize@url \@href}%
\providecommand \@href[1]{\@@startlink{#1}\@@href}%
\providecommand \@@href[1]{\endgroup#1\@@endlink}%
\providecommand \@sanitize@url [0]{\catcode `\\12\catcode `\$12\catcode `\&12\catcode `\#12\catcode `\^12\catcode `\_12\catcode `\%12\relax}%
\providecommand \@@startlink[1]{}%
\providecommand \@@endlink[0]{}%
\providecommand \url  [0]{\begingroup\@sanitize@url \@url }%
\providecommand \@url [1]{\endgroup\@href {#1}{\urlprefix }}%
\providecommand \urlprefix  [0]{URL }%
\providecommand \Eprint [0]{\href }%
\providecommand \doibase [0]{http://dx.doi.org/}%
\providecommand \selectlanguage [0]{\@gobble}%
\providecommand \bibinfo  [0]{\@secondoftwo}%
\providecommand \bibfield  [0]{\@secondoftwo}%
\providecommand \translation [1]{[#1]}%
\providecommand \BibitemOpen [0]{}%
\providecommand \bibitemStop [0]{}%
\providecommand \bibitemNoStop [0]{.\EOS\space}%
\providecommand \EOS [0]{\spacefactor3000\relax}%
\providecommand \BibitemShut  [1]{\csname bibitem#1\endcsname}%
\let\auto@bib@innerbib\@empty
\bibitem [{\citenamefont {Konoplya}\ and\ \citenamefont {Zhidenko}(2011)}]{Konoplya:2011qq}%
  \BibitemOpen
  \bibfield  {author} {\bibinfo {author} {\bibfnamefont {R.~A.}\ \bibnamefont {Konoplya}}\ and\ \bibinfo {author} {\bibfnamefont {A.}~\bibnamefont {Zhidenko}},\ }\href {\doibase 10.1103/RevModPhys.83.793} {\bibfield  {journal} {\bibinfo  {journal} {Rev. Mod. Phys.}\ }\textbf {\bibinfo {volume} {83}},\ \bibinfo {pages} {793} (\bibinfo {year} {2011})},\ \Eprint {http://arxiv.org/abs/1102.4014} {arXiv:1102.4014 [gr-qc]} \BibitemShut {NoStop}%
\bibitem [{\citenamefont {Kokkotas}\ and\ \citenamefont {Schmidt}(1999)}]{Kokkotas:1999bd}%
  \BibitemOpen
  \bibfield  {author} {\bibinfo {author} {\bibfnamefont {K.~D.}\ \bibnamefont {Kokkotas}}\ and\ \bibinfo {author} {\bibfnamefont {B.~G.}\ \bibnamefont {Schmidt}},\ }\href {\doibase 10.12942/lrr-1999-2} {\bibfield  {journal} {\bibinfo  {journal} {Living Rev. Rel.}\ }\textbf {\bibinfo {volume} {2}},\ \bibinfo {pages} {2} (\bibinfo {year} {1999})},\ \Eprint {http://arxiv.org/abs/gr-qc/9909058} {arXiv:gr-qc/9909058} \BibitemShut {NoStop}%
\bibitem [{\citenamefont {Nollert}(1999)}]{Nollert:1999ji}%
  \BibitemOpen
  \bibfield  {author} {\bibinfo {author} {\bibfnamefont {H.-P.}\ \bibnamefont {Nollert}},\ }\href {\doibase 10.1088/0264-9381/16/12/201} {\bibfield  {journal} {\bibinfo  {journal} {Class. Quant. Grav.}\ }\textbf {\bibinfo {volume} {16}},\ \bibinfo {pages} {R159} (\bibinfo {year} {1999})}\BibitemShut {NoStop}%
\bibitem [{\citenamefont {Bolokhov}\ and\ \citenamefont {Skvortsova}(2025{\natexlab{a}})}]{Bolokhov:2025uxz}%
  \BibitemOpen
  \bibfield  {author} {\bibinfo {author} {\bibfnamefont {S.~V.}\ \bibnamefont {Bolokhov}}\ and\ \bibinfo {author} {\bibfnamefont {M.}~\bibnamefont {Skvortsova}},\ }\href {\doibase 10.1134/S0202289325700306} {\bibfield  {journal} {\bibinfo  {journal} {Grav. Cosmol.}\ }\textbf {\bibinfo {volume} {31}},\ \bibinfo {pages} {423} (\bibinfo {year} {2025}{\natexlab{a}})},\ \Eprint {http://arxiv.org/abs/2504.05014} {arXiv:2504.05014 [gr-qc]} \BibitemShut {NoStop}%
\bibitem [{\citenamefont {Hawking}(1975)}]{Hawking:1975vcx}%
  \BibitemOpen
  \bibfield  {author} {\bibinfo {author} {\bibfnamefont {S.~W.}\ \bibnamefont {Hawking}},\ }\href {\doibase 10.1007/BF02345020} {\bibfield  {journal} {\bibinfo  {journal} {Commun. Math. Phys.}\ }\textbf {\bibinfo {volume} {43}},\ \bibinfo {pages} {199} (\bibinfo {year} {1975})},\ \bibinfo {note} {[Erratum: Commun.Math.Phys. 46, 206 (1976)]}\BibitemShut {NoStop}%
\bibitem [{\citenamefont {Page}(1976{\natexlab{a}})}]{Page:1976df}%
  \BibitemOpen
  \bibfield  {author} {\bibinfo {author} {\bibfnamefont {D.~N.}\ \bibnamefont {Page}},\ }\href {\doibase 10.1103/PhysRevD.13.198} {\bibfield  {journal} {\bibinfo  {journal} {Phys. Rev. D}\ }\textbf {\bibinfo {volume} {13}},\ \bibinfo {pages} {198} (\bibinfo {year} {1976}{\natexlab{a}})}\BibitemShut {NoStop}%
\bibitem [{\citenamefont {Page}(1976{\natexlab{b}})}]{Page:1976ki}%
  \BibitemOpen
  \bibfield  {author} {\bibinfo {author} {\bibfnamefont {D.~N.}\ \bibnamefont {Page}},\ }\href {\doibase 10.1103/PhysRevD.14.3260} {\bibfield  {journal} {\bibinfo  {journal} {Phys. Rev. D}\ }\textbf {\bibinfo {volume} {14}},\ \bibinfo {pages} {3260} (\bibinfo {year} {1976}{\natexlab{b}})}\BibitemShut {NoStop}%
\bibitem [{\citenamefont {Kanti}(2004)}]{Kanti:2004nr}%
  \BibitemOpen
  \bibfield  {author} {\bibinfo {author} {\bibfnamefont {P.}~\bibnamefont {Kanti}},\ }\href {\doibase 10.1142/S0217751X04018324} {\bibfield  {journal} {\bibinfo  {journal} {Int. J. Mod. Phys. A}\ }\textbf {\bibinfo {volume} {19}},\ \bibinfo {pages} {4899} (\bibinfo {year} {2004})},\ \Eprint {http://arxiv.org/abs/hep-ph/0402168} {arXiv:hep-ph/0402168} \BibitemShut {NoStop}%
\bibitem [{\citenamefont {Konoplya}\ and\ \citenamefont {Zhidenko}(2025)}]{Konoplya:2025uta}%
  \BibitemOpen
  \bibfield  {author} {\bibinfo {author} {\bibfnamefont {R.~A.}\ \bibnamefont {Konoplya}}\ and\ \bibinfo {author} {\bibfnamefont {A.}~\bibnamefont {Zhidenko}},\ }\href {\doibase 10.53941/ijgtp.2025.100005} {\bibfield  {journal} {\bibinfo  {journal} {Int. J. Grav. Theor. Phys.}\ }\textbf {\bibinfo {volume} {1}},\ \bibinfo {pages} {5} (\bibinfo {year} {2025})},\ \Eprint {http://arxiv.org/abs/2507.22660} {arXiv:2507.22660 [gr-qc]} \BibitemShut {NoStop}%
\bibitem [{\citenamefont {Duffy}\ \emph {et~al.}(2005)\citenamefont {Duffy}, \citenamefont {Harris}, \citenamefont {Kanti},\ and\ \citenamefont {Winstanley}}]{Duffy:2005ns}%
  \BibitemOpen
  \bibfield  {author} {\bibinfo {author} {\bibfnamefont {G.}~\bibnamefont {Duffy}}, \bibinfo {author} {\bibfnamefont {C.}~\bibnamefont {Harris}}, \bibinfo {author} {\bibfnamefont {P.}~\bibnamefont {Kanti}}, \ and\ \bibinfo {author} {\bibfnamefont {E.}~\bibnamefont {Winstanley}},\ }\href {\doibase 10.1088/1126-6708/2005/09/049} {\bibfield  {journal} {\bibinfo  {journal} {JHEP}\ }\textbf {\bibinfo {volume} {09}},\ \bibinfo {pages} {049} (\bibinfo {year} {2005})},\ \Eprint {http://arxiv.org/abs/hep-th/0507274} {arXiv:hep-th/0507274} \BibitemShut {NoStop}%
\bibitem [{\citenamefont {Kanti}\ and\ \citenamefont {March-Russell}(2003)}]{Kanti:2002ge}%
  \BibitemOpen
  \bibfield  {author} {\bibinfo {author} {\bibfnamefont {P.}~\bibnamefont {Kanti}}\ and\ \bibinfo {author} {\bibfnamefont {J.}~\bibnamefont {March-Russell}},\ }\href {\doibase 10.1103/PhysRevD.67.104019} {\bibfield  {journal} {\bibinfo  {journal} {Phys. Rev. D}\ }\textbf {\bibinfo {volume} {67}},\ \bibinfo {pages} {104019} (\bibinfo {year} {2003})},\ \Eprint {http://arxiv.org/abs/hep-ph/0212199} {arXiv:hep-ph/0212199} \BibitemShut {NoStop}%
\bibitem [{\citenamefont {Klebanov}\ and\ \citenamefont {Mathur}(1997)}]{Klebanov:1997cx}%
  \BibitemOpen
  \bibfield  {author} {\bibinfo {author} {\bibfnamefont {I.~R.}\ \bibnamefont {Klebanov}}\ and\ \bibinfo {author} {\bibfnamefont {S.~D.}\ \bibnamefont {Mathur}},\ }\href {\doibase 10.1016/S0550-3213(97)00287-3} {\bibfield  {journal} {\bibinfo  {journal} {Nucl. Phys. B}\ }\textbf {\bibinfo {volume} {500}},\ \bibinfo {pages} {115} (\bibinfo {year} {1997})},\ \Eprint {http://arxiv.org/abs/hep-th/9701187} {arXiv:hep-th/9701187} \BibitemShut {NoStop}%
\bibitem [{\citenamefont {Konoplya}(2023{\natexlab{a}})}]{Konoplya:2023ppx}%
  \BibitemOpen
  \bibfield  {author} {\bibinfo {author} {\bibfnamefont {R.~A.}\ \bibnamefont {Konoplya}},\ }\href {\doibase 10.1088/1475-7516/2023/07/001} {\bibfield  {journal} {\bibinfo  {journal} {JCAP}\ }\textbf {\bibinfo {volume} {07}},\ \bibinfo {pages} {001} (\bibinfo {year} {2023}{\natexlab{a}})},\ \Eprint {http://arxiv.org/abs/2305.09187} {arXiv:2305.09187 [gr-qc]} \BibitemShut {NoStop}%
\bibitem [{\citenamefont {Pappas}\ \emph {et~al.}(2016)\citenamefont {Pappas}, \citenamefont {Kanti},\ and\ \citenamefont {Pappas}}]{Pappas:2016ovo}%
  \BibitemOpen
  \bibfield  {author} {\bibinfo {author} {\bibfnamefont {T.}~\bibnamefont {Pappas}}, \bibinfo {author} {\bibfnamefont {P.}~\bibnamefont {Kanti}}, \ and\ \bibinfo {author} {\bibfnamefont {N.}~\bibnamefont {Pappas}},\ }\href {\doibase 10.1103/PhysRevD.94.024035} {\bibfield  {journal} {\bibinfo  {journal} {Phys. Rev. D}\ }\textbf {\bibinfo {volume} {94}},\ \bibinfo {pages} {024035} (\bibinfo {year} {2016})},\ \Eprint {http://arxiv.org/abs/1604.08617} {arXiv:1604.08617 [hep-th]} \BibitemShut {NoStop}%
\bibitem [{\citenamefont {Harris}\ and\ \citenamefont {Kanti}(2003)}]{Harris:2003eg}%
  \BibitemOpen
  \bibfield  {author} {\bibinfo {author} {\bibfnamefont {C.~M.}\ \bibnamefont {Harris}}\ and\ \bibinfo {author} {\bibfnamefont {P.}~\bibnamefont {Kanti}},\ }\href {\doibase 10.1088/1126-6708/2003/10/014} {\bibfield  {journal} {\bibinfo  {journal} {JHEP}\ }\textbf {\bibinfo {volume} {10}},\ \bibinfo {pages} {014} (\bibinfo {year} {2003})},\ \Eprint {http://arxiv.org/abs/hep-ph/0309054} {arXiv:hep-ph/0309054} \BibitemShut {NoStop}%
\bibitem [{\citenamefont {Bolokhov}\ and\ \citenamefont {Konoplya}(2025)}]{Bolokhov:2024voa}%
  \BibitemOpen
  \bibfield  {author} {\bibinfo {author} {\bibfnamefont {S.~V.}\ \bibnamefont {Bolokhov}}\ and\ \bibinfo {author} {\bibfnamefont {R.~A.}\ \bibnamefont {Konoplya}},\ }\href {\doibase 10.1103/PhysRevD.111.064007} {\bibfield  {journal} {\bibinfo  {journal} {Phys. Rev. D}\ }\textbf {\bibinfo {volume} {111}},\ \bibinfo {pages} {064007} (\bibinfo {year} {2025})},\ \Eprint {http://arxiv.org/abs/2410.10419} {arXiv:2410.10419 [gr-qc]} \BibitemShut {NoStop}%
\bibitem [{\citenamefont {Konoplya}\ and\ \citenamefont {Zinhailo}(2020)}]{Konoplya:2020cbv}%
  \BibitemOpen
  \bibfield  {author} {\bibinfo {author} {\bibfnamefont {R.~A.}\ \bibnamefont {Konoplya}}\ and\ \bibinfo {author} {\bibfnamefont {A.~F.}\ \bibnamefont {Zinhailo}},\ }\href {\doibase 10.1016/j.physletb.2020.135793} {\bibfield  {journal} {\bibinfo  {journal} {Phys. Lett. B}\ }\textbf {\bibinfo {volume} {810}},\ \bibinfo {pages} {135793} (\bibinfo {year} {2020})},\ \Eprint {http://arxiv.org/abs/2004.02248} {arXiv:2004.02248 [gr-qc]} \BibitemShut {NoStop}%
\bibitem [{\citenamefont {Cvetic}\ and\ \citenamefont {Larsen}(1997{\natexlab{a}})}]{Cvetic:1997uw}%
  \BibitemOpen
  \bibfield  {author} {\bibinfo {author} {\bibfnamefont {M.}~\bibnamefont {Cvetic}}\ and\ \bibinfo {author} {\bibfnamefont {F.}~\bibnamefont {Larsen}},\ }\href {\doibase 10.1103/PhysRevD.56.4994} {\bibfield  {journal} {\bibinfo  {journal} {Phys. Rev. D}\ }\textbf {\bibinfo {volume} {56}},\ \bibinfo {pages} {4994} (\bibinfo {year} {1997}{\natexlab{a}})},\ \Eprint {http://arxiv.org/abs/hep-th/9705192} {arXiv:hep-th/9705192} \BibitemShut {NoStop}%
\bibitem [{\citenamefont {Bonanno}\ \emph {et~al.}(2025)\citenamefont {Bonanno}, \citenamefont {Konoplya}, \citenamefont {Oglialoro},\ and\ \citenamefont {Spina}}]{Bonanno:2025dry}%
  \BibitemOpen
  \bibfield  {author} {\bibinfo {author} {\bibfnamefont {A.~M.}\ \bibnamefont {Bonanno}}, \bibinfo {author} {\bibfnamefont {R.~A.}\ \bibnamefont {Konoplya}}, \bibinfo {author} {\bibfnamefont {G.}~\bibnamefont {Oglialoro}}, \ and\ \bibinfo {author} {\bibfnamefont {A.}~\bibnamefont {Spina}},\ }\href {\doibase 10.1088/1475-7516/2025/12/042} {\bibfield  {journal} {\bibinfo  {journal} {JCAP}\ }\textbf {\bibinfo {volume} {12}},\ \bibinfo {pages} {042} (\bibinfo {year} {2025})},\ \Eprint {http://arxiv.org/abs/2509.12469} {arXiv:2509.12469 [gr-qc]} \BibitemShut {NoStop}%
\bibitem [{\citenamefont {Kanti}\ and\ \citenamefont {Pappas}(2017)}]{Kanti:2017ubd}%
  \BibitemOpen
  \bibfield  {author} {\bibinfo {author} {\bibfnamefont {P.}~\bibnamefont {Kanti}}\ and\ \bibinfo {author} {\bibfnamefont {T.}~\bibnamefont {Pappas}},\ }\href {\doibase 10.1103/PhysRevD.96.024038} {\bibfield  {journal} {\bibinfo  {journal} {Phys. Rev. D}\ }\textbf {\bibinfo {volume} {96}},\ \bibinfo {pages} {024038} (\bibinfo {year} {2017})},\ \Eprint {http://arxiv.org/abs/1705.09108} {arXiv:1705.09108 [hep-th]} \BibitemShut {NoStop}%
\bibitem [{\citenamefont {Gubser}\ and\ \citenamefont {Klebanov}(1996)}]{Gubser:1996zp}%
  \BibitemOpen
  \bibfield  {author} {\bibinfo {author} {\bibfnamefont {S.~S.}\ \bibnamefont {Gubser}}\ and\ \bibinfo {author} {\bibfnamefont {I.~R.}\ \bibnamefont {Klebanov}},\ }\href {\doibase 10.1103/PhysRevLett.77.4491} {\bibfield  {journal} {\bibinfo  {journal} {Phys. Rev. Lett.}\ }\textbf {\bibinfo {volume} {77}},\ \bibinfo {pages} {4491} (\bibinfo {year} {1996})},\ \Eprint {http://arxiv.org/abs/hep-th/9609076} {arXiv:hep-th/9609076} \BibitemShut {NoStop}%
\bibitem [{\citenamefont {Konoplya}\ and\ \citenamefont {Zhidenko}(2023{\natexlab{a}})}]{Konoplya:2023moy}%
  \BibitemOpen
  \bibfield  {author} {\bibinfo {author} {\bibfnamefont {R.~A.}\ \bibnamefont {Konoplya}}\ and\ \bibinfo {author} {\bibfnamefont {A.}~\bibnamefont {Zhidenko}},\ }\href {\doibase 10.1088/1361-6382/ad0a52} {\bibfield  {journal} {\bibinfo  {journal} {Class. Quant. Grav.}\ }\textbf {\bibinfo {volume} {40}},\ \bibinfo {pages} {245005} (\bibinfo {year} {2023}{\natexlab{a}})},\ \Eprint {http://arxiv.org/abs/2309.02560} {arXiv:2309.02560 [gr-qc]} \BibitemShut {NoStop}%
\bibitem [{\citenamefont {Konoplya}\ \emph {et~al.}(2020)\citenamefont {Konoplya}, \citenamefont {Zinhailo},\ and\ \citenamefont {Stuchlik}}]{Konoplya:2020jgt}%
  \BibitemOpen
  \bibfield  {author} {\bibinfo {author} {\bibfnamefont {R.~A.}\ \bibnamefont {Konoplya}}, \bibinfo {author} {\bibfnamefont {A.~F.}\ \bibnamefont {Zinhailo}}, \ and\ \bibinfo {author} {\bibfnamefont {Z.}~\bibnamefont {Stuchlik}},\ }\href {\doibase 10.1103/PhysRevD.102.044023} {\bibfield  {journal} {\bibinfo  {journal} {Phys. Rev. D}\ }\textbf {\bibinfo {volume} {102}},\ \bibinfo {pages} {044023} (\bibinfo {year} {2020})},\ \Eprint {http://arxiv.org/abs/2006.10462} {arXiv:2006.10462 [gr-qc]} \BibitemShut {NoStop}%
\bibitem [{\citenamefont {Dubinsky}\ and\ \citenamefont {Zinhailo}(2025)}]{Dubinsky:2024nzo}%
  \BibitemOpen
  \bibfield  {author} {\bibinfo {author} {\bibfnamefont {A.}~\bibnamefont {Dubinsky}}\ and\ \bibinfo {author} {\bibfnamefont {A.~F.}\ \bibnamefont {Zinhailo}},\ }\href {\doibase 10.1209/0295-5075/adbc17} {\bibfield  {journal} {\bibinfo  {journal} {EPL}\ }\textbf {\bibinfo {volume} {149}},\ \bibinfo {pages} {69004} (\bibinfo {year} {2025})},\ \Eprint {http://arxiv.org/abs/2410.15232} {arXiv:2410.15232 [gr-qc]} \BibitemShut {NoStop}%
\bibitem [{\citenamefont {Fernando}(2017)}]{Fernando:2016ksb}%
  \BibitemOpen
  \bibfield  {author} {\bibinfo {author} {\bibfnamefont {S.}~\bibnamefont {Fernando}},\ }\href {\doibase 10.1142/S0218271817500717} {\bibfield  {journal} {\bibinfo  {journal} {Int. J. Mod. Phys. D}\ }\textbf {\bibinfo {volume} {26}},\ \bibinfo {pages} {1750071} (\bibinfo {year} {2017})},\ \Eprint {http://arxiv.org/abs/1611.05337} {arXiv:1611.05337 [gr-qc]} \BibitemShut {NoStop}%
\bibitem [{\citenamefont {Malik}(2024{\natexlab{a}})}]{Malik:2024wvs}%
  \BibitemOpen
  \bibfield  {author} {\bibinfo {author} {\bibfnamefont {Z.}~\bibnamefont {Malik}},\ }\href@noop {} {\  (\bibinfo {year} {2024}{\natexlab{a}})},\ \Eprint {http://arxiv.org/abs/2412.13385} {arXiv:2412.13385 [gr-qc]} \BibitemShut {NoStop}%
\bibitem [{\citenamefont {Stashko}(2024)}]{Stashko:2024wuq}%
  \BibitemOpen
  \bibfield  {author} {\bibinfo {author} {\bibfnamefont {O.}~\bibnamefont {Stashko}},\ }\href {\doibase 10.1103/PhysRevD.110.084016} {\bibfield  {journal} {\bibinfo  {journal} {Phys. Rev. D}\ }\textbf {\bibinfo {volume} {110}},\ \bibinfo {pages} {084016} (\bibinfo {year} {2024})},\ \Eprint {http://arxiv.org/abs/2407.07892} {arXiv:2407.07892 [gr-qc]} \BibitemShut {NoStop}%
\bibitem [{\citenamefont {Konoplya}(2021)}]{Konoplya:2021ube}%
  \BibitemOpen
  \bibfield  {author} {\bibinfo {author} {\bibfnamefont {R.~A.}\ \bibnamefont {Konoplya}},\ }\href {\doibase 10.1016/j.physletb.2021.136734} {\bibfield  {journal} {\bibinfo  {journal} {Phys. Lett. B}\ }\textbf {\bibinfo {volume} {823}},\ \bibinfo {pages} {136734} (\bibinfo {year} {2021})},\ \Eprint {http://arxiv.org/abs/2109.01640} {arXiv:2109.01640 [gr-qc]} \BibitemShut {NoStop}%
\bibitem [{\citenamefont {L{\"u}tf{\"u}o{\u{g}}lu}(2025{\natexlab{a}})}]{Lutfuoglu:2025ljm}%
  \BibitemOpen
  \bibfield  {author} {\bibinfo {author} {\bibfnamefont {B.~C.}\ \bibnamefont {L{\"u}tf{\"u}o{\u{g}}lu}},\ }\href {\doibase 10.1140/epjc/s10052-025-14380-x} {\bibfield  {journal} {\bibinfo  {journal} {Eur. Phys. J. C}\ }\textbf {\bibinfo {volume} {85}},\ \bibinfo {pages} {630} (\bibinfo {year} {2025}{\natexlab{a}})},\ \Eprint {http://arxiv.org/abs/2504.18482} {arXiv:2504.18482 [gr-qc]} \BibitemShut {NoStop}%
\bibitem [{\citenamefont {Konoplya}\ \emph {et~al.}(2023)\citenamefont {Konoplya}, \citenamefont {Ovchinnikov},\ and\ \citenamefont {Ahmedov}}]{Konoplya:2023ahd}%
  \BibitemOpen
  \bibfield  {author} {\bibinfo {author} {\bibfnamefont {R.~A.}\ \bibnamefont {Konoplya}}, \bibinfo {author} {\bibfnamefont {D.}~\bibnamefont {Ovchinnikov}}, \ and\ \bibinfo {author} {\bibfnamefont {B.}~\bibnamefont {Ahmedov}},\ }\href {\doibase 10.1103/PhysRevD.108.104054} {\bibfield  {journal} {\bibinfo  {journal} {Phys. Rev. D}\ }\textbf {\bibinfo {volume} {108}},\ \bibinfo {pages} {104054} (\bibinfo {year} {2023})},\ \Eprint {http://arxiv.org/abs/2307.10801} {arXiv:2307.10801 [gr-qc]} \BibitemShut {NoStop}%
\bibitem [{\citenamefont {Harris}\ and\ \citenamefont {Kanti}(2006)}]{Harris:2005jx}%
  \BibitemOpen
  \bibfield  {author} {\bibinfo {author} {\bibfnamefont {C.~M.}\ \bibnamefont {Harris}}\ and\ \bibinfo {author} {\bibfnamefont {P.}~\bibnamefont {Kanti}},\ }\href {\doibase 10.1016/j.physletb.2005.10.025} {\bibfield  {journal} {\bibinfo  {journal} {Phys. Lett. B}\ }\textbf {\bibinfo {volume} {633}},\ \bibinfo {pages} {106} (\bibinfo {year} {2006})},\ \Eprint {http://arxiv.org/abs/hep-th/0503010} {arXiv:hep-th/0503010} \BibitemShut {NoStop}%
\bibitem [{\citenamefont {Pappas}\ and\ \citenamefont {Kanti}(2017)}]{Pappas:2017kam}%
  \BibitemOpen
  \bibfield  {author} {\bibinfo {author} {\bibfnamefont {T.}~\bibnamefont {Pappas}}\ and\ \bibinfo {author} {\bibfnamefont {P.}~\bibnamefont {Kanti}},\ }\href {\doibase 10.1016/j.physletb.2017.10.058} {\bibfield  {journal} {\bibinfo  {journal} {Phys. Lett. B}\ }\textbf {\bibinfo {volume} {775}},\ \bibinfo {pages} {140} (\bibinfo {year} {2017})},\ \Eprint {http://arxiv.org/abs/1707.04900} {arXiv:1707.04900 [hep-th]} \BibitemShut {NoStop}%
\bibitem [{\citenamefont {Cvetic}\ and\ \citenamefont {Larsen}(1997{\natexlab{b}})}]{Cvetic:1997xv}%
  \BibitemOpen
  \bibfield  {author} {\bibinfo {author} {\bibfnamefont {M.}~\bibnamefont {Cvetic}}\ and\ \bibinfo {author} {\bibfnamefont {F.}~\bibnamefont {Larsen}},\ }\href {\doibase 10.1016/S0550-3213(97)00541-5} {\bibfield  {journal} {\bibinfo  {journal} {Nucl. Phys. B}\ }\textbf {\bibinfo {volume} {506}},\ \bibinfo {pages} {107} (\bibinfo {year} {1997}{\natexlab{b}})},\ \Eprint {http://arxiv.org/abs/hep-th/9706071} {arXiv:hep-th/9706071} \BibitemShut {NoStop}%
\bibitem [{\citenamefont {Bolokhov}\ \emph {et~al.}(2025)\citenamefont {Bolokhov}, \citenamefont {Bronnikov},\ and\ \citenamefont {Konoplya}}]{Bolokhov:2023ozp}%
  \BibitemOpen
  \bibfield  {author} {\bibinfo {author} {\bibfnamefont {S.}~\bibnamefont {Bolokhov}}, \bibinfo {author} {\bibfnamefont {K.}~\bibnamefont {Bronnikov}}, \ and\ \bibinfo {author} {\bibfnamefont {R.}~\bibnamefont {Konoplya}},\ }\href {\doibase 10.1002/prop.202400187} {\bibfield  {journal} {\bibinfo  {journal} {Fortsch. Phys.}\ }\textbf {\bibinfo {volume} {73}},\ \bibinfo {pages} {2400187} (\bibinfo {year} {2025})},\ \Eprint {http://arxiv.org/abs/2306.11083} {arXiv:2306.11083 [gr-qc]} \BibitemShut {NoStop}%
\bibitem [{\citenamefont {Kokkotas}\ \emph {et~al.}(2011)\citenamefont {Kokkotas}, \citenamefont {Konoplya},\ and\ \citenamefont {Zhidenko}}]{Kokkotas:2010zd}%
  \BibitemOpen
  \bibfield  {author} {\bibinfo {author} {\bibfnamefont {K.~D.}\ \bibnamefont {Kokkotas}}, \bibinfo {author} {\bibfnamefont {R.~A.}\ \bibnamefont {Konoplya}}, \ and\ \bibinfo {author} {\bibfnamefont {A.}~\bibnamefont {Zhidenko}},\ }\href {\doibase 10.1103/PhysRevD.83.024031} {\bibfield  {journal} {\bibinfo  {journal} {Phys. Rev. D}\ }\textbf {\bibinfo {volume} {83}},\ \bibinfo {pages} {024031} (\bibinfo {year} {2011})},\ \Eprint {http://arxiv.org/abs/1011.1843} {arXiv:1011.1843 [gr-qc]} \BibitemShut {NoStop}%
\bibitem [{\citenamefont {Kanti}\ and\ \citenamefont {Winstanley}(2015)}]{Kanti:2014vsa}%
  \BibitemOpen
  \bibfield  {author} {\bibinfo {author} {\bibfnamefont {P.}~\bibnamefont {Kanti}}\ and\ \bibinfo {author} {\bibfnamefont {E.}~\bibnamefont {Winstanley}},\ }\href {\doibase 10.1007/978-3-319-10852-0_8} {\bibfield  {journal} {\bibinfo  {journal} {Fundam. Theor. Phys.}\ }\textbf {\bibinfo {volume} {178}},\ \bibinfo {pages} {229} (\bibinfo {year} {2015})},\ \Eprint {http://arxiv.org/abs/1402.3952} {arXiv:1402.3952 [hep-th]} \BibitemShut {NoStop}%
\bibitem [{\citenamefont {Konoplya}\ and\ \citenamefont {Zinhailo}(2019)}]{Konoplya:2019ppy}%
  \BibitemOpen
  \bibfield  {author} {\bibinfo {author} {\bibfnamefont {R.~A.}\ \bibnamefont {Konoplya}}\ and\ \bibinfo {author} {\bibfnamefont {A.~F.}\ \bibnamefont {Zinhailo}},\ }\href {\doibase 10.1103/PhysRevD.99.104060} {\bibfield  {journal} {\bibinfo  {journal} {Phys. Rev. D}\ }\textbf {\bibinfo {volume} {99}},\ \bibinfo {pages} {104060} (\bibinfo {year} {2019})},\ \Eprint {http://arxiv.org/abs/1904.05341} {arXiv:1904.05341 [gr-qc]} \BibitemShut {NoStop}%
\bibitem [{\citenamefont {Konoplya}(2020)}]{Konoplya:2019xmn}%
  \BibitemOpen
  \bibfield  {author} {\bibinfo {author} {\bibfnamefont {R.~A.}\ \bibnamefont {Konoplya}},\ }\href {\doibase 10.1016/j.physletb.2020.135363} {\bibfield  {journal} {\bibinfo  {journal} {Phys. Lett. B}\ }\textbf {\bibinfo {volume} {804}},\ \bibinfo {pages} {135363} (\bibinfo {year} {2020})},\ \Eprint {http://arxiv.org/abs/1912.10582} {arXiv:1912.10582 [gr-qc]} \BibitemShut {NoStop}%
\bibitem [{\citenamefont {Konoplya}(2002)}]{Konoplya:2001ji}%
  \BibitemOpen
  \bibfield  {author} {\bibinfo {author} {\bibfnamefont {R.~A.}\ \bibnamefont {Konoplya}},\ }\href {\doibase 10.1023/A:1015347628961} {\bibfield  {journal} {\bibinfo  {journal} {Gen. Rel. Grav.}\ }\textbf {\bibinfo {volume} {34}},\ \bibinfo {pages} {329} (\bibinfo {year} {2002})},\ \Eprint {http://arxiv.org/abs/gr-qc/0109096} {arXiv:gr-qc/0109096} \BibitemShut {NoStop}%
\bibitem [{\citenamefont {Konoplya}\ and\ \citenamefont {Zhidenko}(2023{\natexlab{b}})}]{Konoplya:2022zav}%
  \BibitemOpen
  \bibfield  {author} {\bibinfo {author} {\bibfnamefont {R.~A.}\ \bibnamefont {Konoplya}}\ and\ \bibinfo {author} {\bibfnamefont {A.}~\bibnamefont {Zhidenko}},\ }\href {\doibase 10.1103/PhysRevD.107.044009} {\bibfield  {journal} {\bibinfo  {journal} {Phys. Rev. D}\ }\textbf {\bibinfo {volume} {107}},\ \bibinfo {pages} {044009} (\bibinfo {year} {2023}{\natexlab{b}})},\ \Eprint {http://arxiv.org/abs/2211.02997} {arXiv:2211.02997 [gr-qc]} \BibitemShut {NoStop}%
\bibitem [{\citenamefont {Fernando}(2005)}]{Fernando:2004ay}%
  \BibitemOpen
  \bibfield  {author} {\bibinfo {author} {\bibfnamefont {S.}~\bibnamefont {Fernando}},\ }\href {\doibase 10.1007/s10714-005-0035-x} {\bibfield  {journal} {\bibinfo  {journal} {Gen. Rel. Grav.}\ }\textbf {\bibinfo {volume} {37}},\ \bibinfo {pages} {461} (\bibinfo {year} {2005})},\ \Eprint {http://arxiv.org/abs/hep-th/0407163} {arXiv:hep-th/0407163} \BibitemShut {NoStop}%
\bibitem [{\citenamefont {Oshita}(2024)}]{Oshita:2023cjz}%
  \BibitemOpen
  \bibfield  {author} {\bibinfo {author} {\bibfnamefont {N.}~\bibnamefont {Oshita}},\ }\href {\doibase 10.1103/PhysRevD.109.104028} {\bibfield  {journal} {\bibinfo  {journal} {Phys. Rev. D}\ }\textbf {\bibinfo {volume} {109}},\ \bibinfo {pages} {104028} (\bibinfo {year} {2024})},\ \Eprint {http://arxiv.org/abs/2309.05725} {arXiv:2309.05725 [gr-qc]} \BibitemShut {NoStop}%
\bibitem [{\citenamefont {Konoplya}\ and\ \citenamefont {Zhidenko}(2024)}]{Konoplya:2024lir}%
  \BibitemOpen
  \bibfield  {author} {\bibinfo {author} {\bibfnamefont {R.~A.}\ \bibnamefont {Konoplya}}\ and\ \bibinfo {author} {\bibfnamefont {A.}~\bibnamefont {Zhidenko}},\ }\href {\doibase 10.1088/1475-7516/2024/09/068} {\bibfield  {journal} {\bibinfo  {journal} {JCAP}\ }\textbf {\bibinfo {volume} {09}},\ \bibinfo {pages} {068} (\bibinfo {year} {2024})},\ \Eprint {http://arxiv.org/abs/2406.11694} {arXiv:2406.11694 [gr-qc]} \BibitemShut {NoStop}%
\bibitem [{\citenamefont {Abbott}\ \emph {et~al.}(2016)\citenamefont {Abbott} \emph {et~al.}}]{LIGOScientific:2016aoc}%
  \BibitemOpen
  \bibfield  {author} {\bibinfo {author} {\bibfnamefont {B.~P.}\ \bibnamefont {Abbott}} \emph {et~al.} (\bibinfo {collaboration} {LIGO Scientific, Virgo}),\ }\href {\doibase 10.1103/PhysRevLett.116.061102} {\bibfield  {journal} {\bibinfo  {journal} {Phys. Rev. Lett.}\ }\textbf {\bibinfo {volume} {116}},\ \bibinfo {pages} {061102} (\bibinfo {year} {2016})},\ \Eprint {http://arxiv.org/abs/1602.03837} {arXiv:1602.03837 [gr-qc]} \BibitemShut {NoStop}%
\bibitem [{\citenamefont {Abbott}\ \emph {et~al.}(2017)\citenamefont {Abbott} \emph {et~al.}}]{LIGOScientific:2017vwq}%
  \BibitemOpen
  \bibfield  {author} {\bibinfo {author} {\bibfnamefont {B.~P.}\ \bibnamefont {Abbott}} \emph {et~al.} (\bibinfo {collaboration} {LIGO Scientific, Virgo}),\ }\href {\doibase 10.1103/PhysRevLett.119.161101} {\bibfield  {journal} {\bibinfo  {journal} {Phys. Rev. Lett.}\ }\textbf {\bibinfo {volume} {119}},\ \bibinfo {pages} {161101} (\bibinfo {year} {2017})},\ \Eprint {http://arxiv.org/abs/1710.05832} {arXiv:1710.05832 [gr-qc]} \BibitemShut {NoStop}%
\bibitem [{\citenamefont {Abbott}\ \emph {et~al.}(2020)\citenamefont {Abbott} \emph {et~al.}}]{LIGOScientific:2020zkf}%
  \BibitemOpen
  \bibfield  {author} {\bibinfo {author} {\bibfnamefont {R.}~\bibnamefont {Abbott}} \emph {et~al.} (\bibinfo {collaboration} {LIGO Scientific, Virgo}),\ }\href {\doibase 10.3847/2041-8213/ab960f} {\bibfield  {journal} {\bibinfo  {journal} {Astrophys. J. Lett.}\ }\textbf {\bibinfo {volume} {896}},\ \bibinfo {pages} {L44} (\bibinfo {year} {2020})},\ \Eprint {http://arxiv.org/abs/2006.12611} {arXiv:2006.12611 [astro-ph.HE]} \BibitemShut {NoStop}%
\bibitem [{\citenamefont {Babak}\ \emph {et~al.}(2017)\citenamefont {Babak}, \citenamefont {Gair}, \citenamefont {Sesana}, \citenamefont {Barausse}, \citenamefont {Sopuerta}, \citenamefont {Berry}, \citenamefont {Berti}, \citenamefont {Amaro-Seoane}, \citenamefont {Petiteau},\ and\ \citenamefont {Klein}}]{Babak:2017tow}%
  \BibitemOpen
  \bibfield  {author} {\bibinfo {author} {\bibfnamefont {S.}~\bibnamefont {Babak}}, \bibinfo {author} {\bibfnamefont {J.}~\bibnamefont {Gair}}, \bibinfo {author} {\bibfnamefont {A.}~\bibnamefont {Sesana}}, \bibinfo {author} {\bibfnamefont {E.}~\bibnamefont {Barausse}}, \bibinfo {author} {\bibfnamefont {C.~F.}\ \bibnamefont {Sopuerta}}, \bibinfo {author} {\bibfnamefont {C.~P.~L.}\ \bibnamefont {Berry}}, \bibinfo {author} {\bibfnamefont {E.}~\bibnamefont {Berti}}, \bibinfo {author} {\bibfnamefont {P.}~\bibnamefont {Amaro-Seoane}}, \bibinfo {author} {\bibfnamefont {A.}~\bibnamefont {Petiteau}}, \ and\ \bibinfo {author} {\bibfnamefont {A.}~\bibnamefont {Klein}},\ }\href {\doibase 10.1103/PhysRevD.95.103012} {\bibfield  {journal} {\bibinfo  {journal} {Phys. Rev. D}\ }\textbf {\bibinfo {volume} {95}},\ \bibinfo {pages} {103012} (\bibinfo {year} {2017})},\ \Eprint {http://arxiv.org/abs/1703.09722} {arXiv:1703.09722 [gr-qc]} \BibitemShut {NoStop}%
\bibitem [{\citenamefont {Dyatlov}(2012)}]{Dyatlov:2011jd}%
  \BibitemOpen
  \bibfield  {author} {\bibinfo {author} {\bibfnamefont {S.}~\bibnamefont {Dyatlov}},\ }\href {\doibase 10.1007/s00023-012-0159-y} {\bibfield  {journal} {\bibinfo  {journal} {Annales Henri Poincare}\ }\textbf {\bibinfo {volume} {13}},\ \bibinfo {pages} {1101} (\bibinfo {year} {2012})},\ \Eprint {http://arxiv.org/abs/1101.1260} {arXiv:1101.1260 [math.AP]} \BibitemShut {NoStop}%
\bibitem [{\citenamefont {Dyatlov}(2011)}]{Dyatlov:2010hq}%
  \BibitemOpen
  \bibfield  {author} {\bibinfo {author} {\bibfnamefont {S.}~\bibnamefont {Dyatlov}},\ }\href {\doibase 10.1007/s00220-011-1286-x} {\bibfield  {journal} {\bibinfo  {journal} {Commun. Math. Phys.}\ }\textbf {\bibinfo {volume} {306}},\ \bibinfo {pages} {119} (\bibinfo {year} {2011})},\ \Eprint {http://arxiv.org/abs/1003.6128} {arXiv:1003.6128 [math.AP]} \BibitemShut {NoStop}%
\bibitem [{\citenamefont {Konoplya}(2024)}]{Konoplya:2024ptj}%
  \BibitemOpen
  \bibfield  {author} {\bibinfo {author} {\bibfnamefont {R.~A.}\ \bibnamefont {Konoplya}},\ }\href {\doibase 10.1103/PhysRevD.109.104018} {\bibfield  {journal} {\bibinfo  {journal} {Phys. Rev. D}\ }\textbf {\bibinfo {volume} {109}},\ \bibinfo {pages} {104018} (\bibinfo {year} {2024})},\ \Eprint {http://arxiv.org/abs/2401.17106} {arXiv:2401.17106 [gr-qc]} \BibitemShut {NoStop}%
\bibitem [{\citenamefont {Fernando}\ and\ \citenamefont {Arnold}(2004)}]{Fernando:2003wc}%
  \BibitemOpen
  \bibfield  {author} {\bibinfo {author} {\bibfnamefont {S.}~\bibnamefont {Fernando}}\ and\ \bibinfo {author} {\bibfnamefont {K.}~\bibnamefont {Arnold}},\ }\href {\doibase 10.1023/B:GERG.0000035953.31652.88} {\bibfield  {journal} {\bibinfo  {journal} {Gen. Rel. Grav.}\ }\textbf {\bibinfo {volume} {36}},\ \bibinfo {pages} {1805} (\bibinfo {year} {2004})},\ \Eprint {http://arxiv.org/abs/hep-th/0312041} {arXiv:hep-th/0312041} \BibitemShut {NoStop}%
\bibitem [{\citenamefont {Chen}\ and\ \citenamefont {Jing}(2005)}]{Chen:2005rm}%
  \BibitemOpen
  \bibfield  {author} {\bibinfo {author} {\bibfnamefont {S.-B.}\ \bibnamefont {Chen}}\ and\ \bibinfo {author} {\bibfnamefont {J.-L.}\ \bibnamefont {Jing}},\ }\href {\doibase 10.1088/0264-9381/22/6/014} {\bibfield  {journal} {\bibinfo  {journal} {Class. Quant. Grav.}\ }\textbf {\bibinfo {volume} {22}},\ \bibinfo {pages} {1129} (\bibinfo {year} {2005})}\BibitemShut {NoStop}%
\bibitem [{\citenamefont {Lopez-Ortega}(2005)}]{Lopez-Ortega:2005obq}%
  \BibitemOpen
  \bibfield  {author} {\bibinfo {author} {\bibfnamefont {A.}~\bibnamefont {Lopez-Ortega}},\ }\href {\doibase 10.1007/s10714-005-0007-1} {\bibfield  {journal} {\bibinfo  {journal} {Gen. Rel. Grav.}\ }\textbf {\bibinfo {volume} {37}},\ \bibinfo {pages} {167} (\bibinfo {year} {2005})}\BibitemShut {NoStop}%
\bibitem [{\citenamefont {Lopez-Ortega}(2009)}]{Lopez-Ortega:2009jpx}%
  \BibitemOpen
  \bibfield  {author} {\bibinfo {author} {\bibfnamefont {A.}~\bibnamefont {Lopez-Ortega}},\ }\href {\doibase 10.1142/S0218271809015199} {\bibfield  {journal} {\bibinfo  {journal} {Int. J. Mod. Phys. D}\ }\textbf {\bibinfo {volume} {18}},\ \bibinfo {pages} {1441} (\bibinfo {year} {2009})},\ \Eprint {http://arxiv.org/abs/0905.0073} {arXiv:0905.0073 [gr-qc]} \BibitemShut {NoStop}%
\bibitem [{\citenamefont {Zinhailo}(2019)}]{Zinhailo:2019rwd}%
  \BibitemOpen
  \bibfield  {author} {\bibinfo {author} {\bibfnamefont {A.~F.}\ \bibnamefont {Zinhailo}},\ }\href {\doibase 10.1140/epjc/s10052-019-7425-9} {\bibfield  {journal} {\bibinfo  {journal} {Eur. Phys. J. C}\ }\textbf {\bibinfo {volume} {79}},\ \bibinfo {pages} {912} (\bibinfo {year} {2019})},\ \Eprint {http://arxiv.org/abs/1909.12664} {arXiv:1909.12664 [gr-qc]} \BibitemShut {NoStop}%
\bibitem [{\citenamefont {Ferrari}\ \emph {et~al.}(2001)\citenamefont {Ferrari}, \citenamefont {Pauri},\ and\ \citenamefont {Piazza}}]{Ferrari:2000ep}%
  \BibitemOpen
  \bibfield  {author} {\bibinfo {author} {\bibfnamefont {V.}~\bibnamefont {Ferrari}}, \bibinfo {author} {\bibfnamefont {M.}~\bibnamefont {Pauri}}, \ and\ \bibinfo {author} {\bibfnamefont {F.}~\bibnamefont {Piazza}},\ }\href {\doibase 10.1103/PhysRevD.63.064009} {\bibfield  {journal} {\bibinfo  {journal} {Phys. Rev. D}\ }\textbf {\bibinfo {volume} {63}},\ \bibinfo {pages} {064009} (\bibinfo {year} {2001})},\ \Eprint {http://arxiv.org/abs/gr-qc/0005125} {arXiv:gr-qc/0005125} \BibitemShut {NoStop}%
\bibitem [{\citenamefont {Carson}\ and\ \citenamefont {Yagi}(2020)}]{Carson:2020ter}%
  \BibitemOpen
  \bibfield  {author} {\bibinfo {author} {\bibfnamefont {Z.}~\bibnamefont {Carson}}\ and\ \bibinfo {author} {\bibfnamefont {K.}~\bibnamefont {Yagi}},\ }\href {\doibase 10.1103/PhysRevD.101.104030} {\bibfield  {journal} {\bibinfo  {journal} {Phys. Rev. D}\ }\textbf {\bibinfo {volume} {101}},\ \bibinfo {pages} {104030} (\bibinfo {year} {2020})},\ \Eprint {http://arxiv.org/abs/2003.00286} {arXiv:2003.00286 [gr-qc]} \BibitemShut {NoStop}%
\bibitem [{\citenamefont {Malybayev}\ \emph {et~al.}(2021)\citenamefont {Malybayev}, \citenamefont {Boshkayev},\ and\ \citenamefont {Ivashchuk}}]{Malybayev:2021lfq}%
  \BibitemOpen
  \bibfield  {author} {\bibinfo {author} {\bibfnamefont {A.~N.}\ \bibnamefont {Malybayev}}, \bibinfo {author} {\bibfnamefont {K.~A.}\ \bibnamefont {Boshkayev}}, \ and\ \bibinfo {author} {\bibfnamefont {V.~D.}\ \bibnamefont {Ivashchuk}},\ }\href {\doibase 10.1140/epjc/s10052-021-09252-z} {\bibfield  {journal} {\bibinfo  {journal} {Eur. Phys. J. C}\ }\textbf {\bibinfo {volume} {81}},\ \bibinfo {pages} {475} (\bibinfo {year} {2021})},\ \Eprint {http://arxiv.org/abs/2103.10920} {arXiv:2103.10920 [gr-qc]} \BibitemShut {NoStop}%
\bibitem [{\citenamefont {Paul}(2024)}]{Paul:2023eep}%
  \BibitemOpen
  \bibfield  {author} {\bibinfo {author} {\bibfnamefont {P.}~\bibnamefont {Paul}},\ }\href {\doibase 10.1140/epjc/s10052-024-12563-6} {\bibfield  {journal} {\bibinfo  {journal} {Eur. Phys. J. C}\ }\textbf {\bibinfo {volume} {84}},\ \bibinfo {pages} {218} (\bibinfo {year} {2024})},\ \Eprint {http://arxiv.org/abs/2312.16479} {arXiv:2312.16479 [gr-qc]} \BibitemShut {NoStop}%
\bibitem [{\citenamefont {Bl\'azquez-Salcedo}\ \emph {et~al.}(2020)\citenamefont {Bl\'azquez-Salcedo}, \citenamefont {Doneva}, \citenamefont {Kahlen}, \citenamefont {Kunz}, \citenamefont {Nedkova},\ and\ \citenamefont {Yazadjiev}}]{Blazquez-Salcedo:2020caw}%
  \BibitemOpen
  \bibfield  {author} {\bibinfo {author} {\bibfnamefont {J.~L.}\ \bibnamefont {Bl\'azquez-Salcedo}}, \bibinfo {author} {\bibfnamefont {D.~D.}\ \bibnamefont {Doneva}}, \bibinfo {author} {\bibfnamefont {S.}~\bibnamefont {Kahlen}}, \bibinfo {author} {\bibfnamefont {J.}~\bibnamefont {Kunz}}, \bibinfo {author} {\bibfnamefont {P.}~\bibnamefont {Nedkova}}, \ and\ \bibinfo {author} {\bibfnamefont {S.~S.}\ \bibnamefont {Yazadjiev}},\ }\href {\doibase 10.1103/PhysRevD.102.024086} {\bibfield  {journal} {\bibinfo  {journal} {Phys. Rev. D}\ }\textbf {\bibinfo {volume} {102}},\ \bibinfo {pages} {024086} (\bibinfo {year} {2020})},\ \Eprint {http://arxiv.org/abs/2006.06006} {arXiv:2006.06006 [gr-qc]} \BibitemShut {NoStop}%
\bibitem [{\citenamefont {Pierini}\ and\ \citenamefont {Gualtieri}(2022)}]{Pierini:2022eim}%
  \BibitemOpen
  \bibfield  {author} {\bibinfo {author} {\bibfnamefont {L.}~\bibnamefont {Pierini}}\ and\ \bibinfo {author} {\bibfnamefont {L.}~\bibnamefont {Gualtieri}},\ }\href {\doibase 10.1103/PhysRevD.106.104009} {\bibfield  {journal} {\bibinfo  {journal} {Phys. Rev. D}\ }\textbf {\bibinfo {volume} {106}},\ \bibinfo {pages} {104009} (\bibinfo {year} {2022})},\ \Eprint {http://arxiv.org/abs/2207.11267} {arXiv:2207.11267 [gr-qc]} \BibitemShut {NoStop}%
\bibitem [{\citenamefont {Dubinsky}\ and\ \citenamefont {Zinhailo}(2024)}]{Dubinsky:2024hmn}%
  \BibitemOpen
  \bibfield  {author} {\bibinfo {author} {\bibfnamefont {A.}~\bibnamefont {Dubinsky}}\ and\ \bibinfo {author} {\bibfnamefont {A.}~\bibnamefont {Zinhailo}},\ }\href {\doibase 10.1140/epjc/s10052-024-13206-6} {\bibfield  {journal} {\bibinfo  {journal} {Eur. Phys. J. C}\ }\textbf {\bibinfo {volume} {84}},\ \bibinfo {pages} {847} (\bibinfo {year} {2024})},\ \Eprint {http://arxiv.org/abs/2404.01834} {arXiv:2404.01834 [gr-qc]} \BibitemShut {NoStop}%
\bibitem [{\citenamefont {Fernando}(2016)}]{Fernando:2016ftj}%
  \BibitemOpen
  \bibfield  {author} {\bibinfo {author} {\bibfnamefont {S.}~\bibnamefont {Fernando}},\ }\href {\doibase 10.1007/s10714-016-2020-y} {\bibfield  {journal} {\bibinfo  {journal} {Gen. Rel. Grav.}\ }\textbf {\bibinfo {volume} {48}},\ \bibinfo {pages} {24} (\bibinfo {year} {2016})},\ \Eprint {http://arxiv.org/abs/1601.06407} {arXiv:1601.06407 [gr-qc]} \BibitemShut {NoStop}%
\bibitem [{\citenamefont {Konoplya}\ and\ \citenamefont {Pappas}(2025)}]{Konoplya:2025ixm}%
  \BibitemOpen
  \bibfield  {author} {\bibinfo {author} {\bibfnamefont {R.~A.}\ \bibnamefont {Konoplya}}\ and\ \bibinfo {author} {\bibfnamefont {T.~D.}\ \bibnamefont {Pappas}},\ }\href@noop {} {\  (\bibinfo {year} {2025})},\ \Eprint {http://arxiv.org/abs/2507.01954} {arXiv:2507.01954 [gr-qc]} \BibitemShut {NoStop}%
\bibitem [{\citenamefont {Konoplya}\ and\ \citenamefont {Stashko}(2025)}]{Konoplya:2025hgp}%
  \BibitemOpen
  \bibfield  {author} {\bibinfo {author} {\bibfnamefont {R.~A.}\ \bibnamefont {Konoplya}}\ and\ \bibinfo {author} {\bibfnamefont {O.~S.}\ \bibnamefont {Stashko}},\ }\href {\doibase 10.1103/PhysRevD.111.084031} {\bibfield  {journal} {\bibinfo  {journal} {Phys. Rev. D}\ }\textbf {\bibinfo {volume} {111}},\ \bibinfo {pages} {084031} (\bibinfo {year} {2025})},\ \Eprint {http://arxiv.org/abs/2502.05689} {arXiv:2502.05689 [gr-qc]} \BibitemShut {NoStop}%
\bibitem [{\citenamefont {Gibbons}\ and\ \citenamefont {Maeda}(1988)}]{Gibbons:1987ps}%
  \BibitemOpen
  \bibfield  {author} {\bibinfo {author} {\bibfnamefont {G.~W.}\ \bibnamefont {Gibbons}}\ and\ \bibinfo {author} {\bibfnamefont {K.-i.}\ \bibnamefont {Maeda}},\ }\href {\doibase 10.1016/0550-3213(88)90006-5} {\bibfield  {journal} {\bibinfo  {journal} {Nucl. Phys. B}\ }\textbf {\bibinfo {volume} {298}},\ \bibinfo {pages} {741} (\bibinfo {year} {1988})}\BibitemShut {NoStop}%
\bibitem [{\citenamefont {Garfinkle}\ \emph {et~al.}(1991)\citenamefont {Garfinkle}, \citenamefont {Horowitz},\ and\ \citenamefont {Strominger}}]{Garfinkle:1990qj}%
  \BibitemOpen
  \bibfield  {author} {\bibinfo {author} {\bibfnamefont {D.}~\bibnamefont {Garfinkle}}, \bibinfo {author} {\bibfnamefont {G.~T.}\ \bibnamefont {Horowitz}}, \ and\ \bibinfo {author} {\bibfnamefont {A.}~\bibnamefont {Strominger}},\ }\href {\doibase 10.1103/PhysRevD.43.3140} {\bibfield  {journal} {\bibinfo  {journal} {Phys. Rev. D}\ }\textbf {\bibinfo {volume} {43}},\ \bibinfo {pages} {3140} (\bibinfo {year} {1991})},\ \bibinfo {note} {[Erratum: Phys.Rev.D 45, 3888 (1992)]}\BibitemShut {NoStop}%
\bibitem [{\citenamefont {Gao}\ and\ \citenamefont {Zhang}(2004)}]{Gao:2004tu}%
  \BibitemOpen
  \bibfield  {author} {\bibinfo {author} {\bibfnamefont {C.~J.}\ \bibnamefont {Gao}}\ and\ \bibinfo {author} {\bibfnamefont {S.~N.}\ \bibnamefont {Zhang}},\ }\href {\doibase 10.1103/PhysRevD.70.124019} {\bibfield  {journal} {\bibinfo  {journal} {Phys. Rev. D}\ }\textbf {\bibinfo {volume} {70}},\ \bibinfo {pages} {124019} (\bibinfo {year} {2004})},\ \Eprint {http://arxiv.org/abs/hep-th/0411104} {arXiv:hep-th/0411104} \BibitemShut {NoStop}%
\bibitem [{\citenamefont {Berti}\ \emph {et~al.}(2009)\citenamefont {Berti}, \citenamefont {Cardoso},\ and\ \citenamefont {Starinets}}]{Berti:2009kk}%
  \BibitemOpen
  \bibfield  {author} {\bibinfo {author} {\bibfnamefont {E.}~\bibnamefont {Berti}}, \bibinfo {author} {\bibfnamefont {V.}~\bibnamefont {Cardoso}}, \ and\ \bibinfo {author} {\bibfnamefont {A.~O.}\ \bibnamefont {Starinets}},\ }\href {\doibase 10.1088/0264-9381/26/16/163001} {\bibfield  {journal} {\bibinfo  {journal} {Class. Quant. Grav.}\ }\textbf {\bibinfo {volume} {26}},\ \bibinfo {pages} {163001} (\bibinfo {year} {2009})},\ \Eprint {http://arxiv.org/abs/0905.2975} {arXiv:0905.2975 [gr-qc]} \BibitemShut {NoStop}%
\bibitem [{\citenamefont {Abdalla}\ \emph {et~al.}(2006)\citenamefont {Abdalla}, \citenamefont {Cuadros-Melgar}, \citenamefont {Pavan},\ and\ \citenamefont {Molina}}]{Abdalla:2006qj}%
  \BibitemOpen
  \bibfield  {author} {\bibinfo {author} {\bibfnamefont {E.}~\bibnamefont {Abdalla}}, \bibinfo {author} {\bibfnamefont {B.}~\bibnamefont {Cuadros-Melgar}}, \bibinfo {author} {\bibfnamefont {A.~B.}\ \bibnamefont {Pavan}}, \ and\ \bibinfo {author} {\bibfnamefont {C.}~\bibnamefont {Molina}},\ }\href {\doibase 10.1016/j.nuclphysb.2006.06.017} {\bibfield  {journal} {\bibinfo  {journal} {Nucl. Phys. B}\ }\textbf {\bibinfo {volume} {752}},\ \bibinfo {pages} {40} (\bibinfo {year} {2006})},\ \Eprint {http://arxiv.org/abs/gr-qc/0604033} {arXiv:gr-qc/0604033} \BibitemShut {NoStop}%
\bibitem [{\citenamefont {Konoplya}\ and\ \citenamefont {Zhidenko}(2007)}]{Konoplya:2006rv}%
  \BibitemOpen
  \bibfield  {author} {\bibinfo {author} {\bibfnamefont {R.~A.}\ \bibnamefont {Konoplya}}\ and\ \bibinfo {author} {\bibfnamefont {A.}~\bibnamefont {Zhidenko}},\ }\href {\doibase 10.1016/j.physletb.2006.11.036} {\bibfield  {journal} {\bibinfo  {journal} {Phys. Lett. B}\ }\textbf {\bibinfo {volume} {644}},\ \bibinfo {pages} {186} (\bibinfo {year} {2007})},\ \Eprint {http://arxiv.org/abs/gr-qc/0605082} {arXiv:gr-qc/0605082} \BibitemShut {NoStop}%
\bibitem [{\citenamefont {Brill}\ and\ \citenamefont {Wheeler}(1957)}]{Brill:1957fx}%
  \BibitemOpen
  \bibfield  {author} {\bibinfo {author} {\bibfnamefont {D.~R.}\ \bibnamefont {Brill}}\ and\ \bibinfo {author} {\bibfnamefont {J.~A.}\ \bibnamefont {Wheeler}},\ }\href {\doibase 10.1103/RevModPhys.29.465} {\bibfield  {journal} {\bibinfo  {journal} {Rev. Mod. Phys.}\ }\textbf {\bibinfo {volume} {29}},\ \bibinfo {pages} {465} (\bibinfo {year} {1957})}\BibitemShut {NoStop}%
\bibitem [{\citenamefont {Schutz}\ and\ \citenamefont {Will}(1985)}]{Schutz:1985km}%
  \BibitemOpen
  \bibfield  {author} {\bibinfo {author} {\bibfnamefont {B.~F.}\ \bibnamefont {Schutz}}\ and\ \bibinfo {author} {\bibfnamefont {C.~M.}\ \bibnamefont {Will}},\ }\href {\doibase 10.1086/184453} {\bibfield  {journal} {\bibinfo  {journal} {Astrophys. J. Lett.}\ }\textbf {\bibinfo {volume} {291}},\ \bibinfo {pages} {L33} (\bibinfo {year} {1985})}\BibitemShut {NoStop}%
\bibitem [{\citenamefont {Iyer}\ and\ \citenamefont {Will}(1987)}]{Iyer:1986np}%
  \BibitemOpen
  \bibfield  {author} {\bibinfo {author} {\bibfnamefont {S.}~\bibnamefont {Iyer}}\ and\ \bibinfo {author} {\bibfnamefont {C.~M.}\ \bibnamefont {Will}},\ }\href {\doibase 10.1103/PhysRevD.35.3621} {\bibfield  {journal} {\bibinfo  {journal} {Phys. Rev. D}\ }\textbf {\bibinfo {volume} {35}},\ \bibinfo {pages} {3621} (\bibinfo {year} {1987})}\BibitemShut {NoStop}%
\bibitem [{\citenamefont {Konoplya}(2003)}]{Konoplya:2003ii}%
  \BibitemOpen
  \bibfield  {author} {\bibinfo {author} {\bibfnamefont {R.~A.}\ \bibnamefont {Konoplya}},\ }\href {\doibase 10.1103/PhysRevD.68.024018} {\bibfield  {journal} {\bibinfo  {journal} {Phys. Rev. D}\ }\textbf {\bibinfo {volume} {68}},\ \bibinfo {pages} {024018} (\bibinfo {year} {2003})},\ \Eprint {http://arxiv.org/abs/gr-qc/0303052} {arXiv:gr-qc/0303052} \BibitemShut {NoStop}%
\bibitem [{\citenamefont {Konoplya}\ \emph {et~al.}(2019)\citenamefont {Konoplya}, \citenamefont {Zhidenko},\ and\ \citenamefont {Zinhailo}}]{Konoplya:2019hlu}%
  \BibitemOpen
  \bibfield  {author} {\bibinfo {author} {\bibfnamefont {R.~A.}\ \bibnamefont {Konoplya}}, \bibinfo {author} {\bibfnamefont {A.}~\bibnamefont {Zhidenko}}, \ and\ \bibinfo {author} {\bibfnamefont {A.~F.}\ \bibnamefont {Zinhailo}},\ }\href {\doibase 10.1088/1361-6382/ab2e25} {\bibfield  {journal} {\bibinfo  {journal} {Class. Quant. Grav.}\ }\textbf {\bibinfo {volume} {36}},\ \bibinfo {pages} {155002} (\bibinfo {year} {2019})},\ \Eprint {http://arxiv.org/abs/1904.10333} {arXiv:1904.10333 [gr-qc]} \BibitemShut {NoStop}%
\bibitem [{\citenamefont {Bolokhov}(2024{\natexlab{a}})}]{Bolokhov:2023ruj}%
  \BibitemOpen
  \bibfield  {author} {\bibinfo {author} {\bibfnamefont {S.~V.}\ \bibnamefont {Bolokhov}},\ }\href {\doibase 10.1103/PhysRevD.109.064017} {\bibfield  {journal} {\bibinfo  {journal} {Phys. Rev. D}\ }\textbf {\bibinfo {volume} {109}},\ \bibinfo {pages} {064017} (\bibinfo {year} {2024}{\natexlab{a}})}\BibitemShut {NoStop}%
\bibitem [{\citenamefont {Bolokhov}(2024{\natexlab{b}})}]{Bolokhov:2023bwm}%
  \BibitemOpen
  \bibfield  {author} {\bibinfo {author} {\bibfnamefont {S.~V.}\ \bibnamefont {Bolokhov}},\ }\href {\doibase 10.1103/PhysRevD.110.024010} {\bibfield  {journal} {\bibinfo  {journal} {Phys. Rev. D}\ }\textbf {\bibinfo {volume} {110}},\ \bibinfo {pages} {024010} (\bibinfo {year} {2024}{\natexlab{b}})},\ \Eprint {http://arxiv.org/abs/2311.05503} {arXiv:2311.05503 [gr-qc]} \BibitemShut {NoStop}%
\bibitem [{\citenamefont {Skvortsova}(2024{\natexlab{a}})}]{Skvortsova:2024wly}%
  \BibitemOpen
  \bibfield  {author} {\bibinfo {author} {\bibfnamefont {M.}~\bibnamefont {Skvortsova}},\ }\href {\doibase 10.1134/S020228932470018X} {\bibfield  {journal} {\bibinfo  {journal} {Grav. Cosmol.}\ }\textbf {\bibinfo {volume} {30}},\ \bibinfo {pages} {279} (\bibinfo {year} {2024}{\natexlab{a}})},\ \Eprint {http://arxiv.org/abs/2405.15807} {arXiv:2405.15807 [gr-qc]} \BibitemShut {NoStop}%
\bibitem [{\citenamefont {Skvortsova}(2025{\natexlab{a}})}]{Skvortsova:2024eqi}%
  \BibitemOpen
  \bibfield  {author} {\bibinfo {author} {\bibfnamefont {M.}~\bibnamefont {Skvortsova}},\ }\href {\doibase 10.1209/0295-5075/adaee2} {\bibfield  {journal} {\bibinfo  {journal} {EPL}\ }\textbf {\bibinfo {volume} {149}},\ \bibinfo {pages} {59001} (\bibinfo {year} {2025}{\natexlab{a}})},\ \Eprint {http://arxiv.org/abs/2503.03650} {arXiv:2503.03650 [gr-qc]} \BibitemShut {NoStop}%
\bibitem [{\citenamefont {Balart}\ \emph {et~al.}(2023)\citenamefont {Balart}, \citenamefont {Panotopoulos},\ and\ \citenamefont {Rinc{\'o}n}}]{Balart:2023odm}%
  \BibitemOpen
  \bibfield  {author} {\bibinfo {author} {\bibfnamefont {L.}~\bibnamefont {Balart}}, \bibinfo {author} {\bibfnamefont {G.}~\bibnamefont {Panotopoulos}}, \ and\ \bibinfo {author} {\bibfnamefont {{\'A}.}~\bibnamefont {Rinc{\'o}n}},\ }\href {\doibase 10.1002/prop.202300075} {\bibfield  {journal} {\bibinfo  {journal} {Fortsch. Phys.}\ }\textbf {\bibinfo {volume} {71}},\ \bibinfo {pages} {2300075} (\bibinfo {year} {2023})},\ \Eprint {http://arxiv.org/abs/2309.01910} {arXiv:2309.01910 [gr-qc]} \BibitemShut {NoStop}%
\bibitem [{\citenamefont {Malik}(2024{\natexlab{b}})}]{Malik:2023bxc}%
  \BibitemOpen
  \bibfield  {author} {\bibinfo {author} {\bibfnamefont {Z.}~\bibnamefont {Malik}},\ }\href {\doibase 10.1007/s10773-024-05737-1} {\bibfield  {journal} {\bibinfo  {journal} {Int. J. Theor. Phys.}\ }\textbf {\bibinfo {volume} {63}},\ \bibinfo {pages} {199} (\bibinfo {year} {2024}{\natexlab{b}})},\ \Eprint {http://arxiv.org/abs/2308.10412} {arXiv:2308.10412 [gr-qc]} \BibitemShut {NoStop}%
\bibitem [{\citenamefont {Malik}(2025{\natexlab{a}})}]{Malik:2025ava}%
  \BibitemOpen
  \bibfield  {author} {\bibinfo {author} {\bibfnamefont {Z.}~\bibnamefont {Malik}},\ }\href {\doibase 10.1016/j.aop.2025.170238} {\bibfield  {journal} {\bibinfo  {journal} {Annals Phys.}\ }\textbf {\bibinfo {volume} {482}},\ \bibinfo {pages} {170238} (\bibinfo {year} {2025}{\natexlab{a}})},\ \Eprint {http://arxiv.org/abs/2504.12570} {arXiv:2504.12570 [gr-qc]} \BibitemShut {NoStop}%
\bibitem [{\citenamefont {Malik}(2025{\natexlab{b}})}]{Malik:2024nhy}%
  \BibitemOpen
  \bibfield  {author} {\bibinfo {author} {\bibfnamefont {Z.}~\bibnamefont {Malik}},\ }\href {\doibase 10.1016/j.aop.2025.170046} {\bibfield  {journal} {\bibinfo  {journal} {Annals Phys.}\ }\textbf {\bibinfo {volume} {479}},\ \bibinfo {pages} {170046} (\bibinfo {year} {2025}{\natexlab{b}})},\ \Eprint {http://arxiv.org/abs/2409.01561} {arXiv:2409.01561 [gr-qc]} \BibitemShut {NoStop}%
\bibitem [{\citenamefont {Al-Badawi}(2023)}]{Al-Badawi:2023lvx}%
  \BibitemOpen
  \bibfield  {author} {\bibinfo {author} {\bibfnamefont {A.}~\bibnamefont {Al-Badawi}},\ }\href {\doibase 10.1140/epjc/s10052-023-11804-4} {\bibfield  {journal} {\bibinfo  {journal} {Eur. Phys. J. C}\ }\textbf {\bibinfo {volume} {83}},\ \bibinfo {pages} {620} (\bibinfo {year} {2023})},\ \Eprint {http://arxiv.org/abs/2307.07974} {arXiv:2307.07974 [gr-qc]} \BibitemShut {NoStop}%
\bibitem [{\citenamefont {Chen}\ and\ \citenamefont {Kotlařík}(2023)}]{Chen:2023akf}%
  \BibitemOpen
  \bibfield  {author} {\bibinfo {author} {\bibfnamefont {C.-Y.}\ \bibnamefont {Chen}}\ and\ \bibinfo {author} {\bibfnamefont {P.}~\bibnamefont {Kotlařík}},\ }\href {\doibase 10.1103/PhysRevD.108.064052} {\bibfield  {journal} {\bibinfo  {journal} {Phys. Rev. D}\ }\textbf {\bibinfo {volume} {108}},\ \bibinfo {pages} {064052} (\bibinfo {year} {2023})},\ \Eprint {http://arxiv.org/abs/2307.07360} {arXiv:2307.07360 [gr-qc]} \BibitemShut {NoStop}%
\bibitem [{\citenamefont {Bolokhov}(2024{\natexlab{c}})}]{Bolokhov:2023dxq}%
  \BibitemOpen
  \bibfield  {author} {\bibinfo {author} {\bibfnamefont {S.~V.}\ \bibnamefont {Bolokhov}},\ }\href {\doibase 10.1016/j.physletb.2024.138879} {\bibfield  {journal} {\bibinfo  {journal} {Phys. Lett. B}\ }\textbf {\bibinfo {volume} {856}},\ \bibinfo {pages} {138879} (\bibinfo {year} {2024}{\natexlab{c}})},\ \Eprint {http://arxiv.org/abs/2310.12326} {arXiv:2310.12326 [gr-qc]} \BibitemShut {NoStop}%
\bibitem [{\citenamefont {Bolokhov}(2024{\natexlab{d}})}]{Bolokhov:2024ixe}%
  \BibitemOpen
  \bibfield  {author} {\bibinfo {author} {\bibfnamefont {S.~V.}\ \bibnamefont {Bolokhov}},\ }\href {\doibase 10.1140/epjc/s10052-024-12990-5} {\bibfield  {journal} {\bibinfo  {journal} {Eur. Phys. J. C}\ }\textbf {\bibinfo {volume} {84}},\ \bibinfo {pages} {634} (\bibinfo {year} {2024}{\natexlab{d}})},\ \Eprint {http://arxiv.org/abs/2404.09364} {arXiv:2404.09364 [gr-qc]} \BibitemShut {NoStop}%
\bibitem [{\citenamefont {Skvortsova}(2024{\natexlab{b}})}]{Skvortsova:2023zmj}%
  \BibitemOpen
  \bibfield  {author} {\bibinfo {author} {\bibfnamefont {M.}~\bibnamefont {Skvortsova}},\ }\href {\doibase 10.1002/prop.202400036} {\bibfield  {journal} {\bibinfo  {journal} {Fortsch. Phys.}\ }\textbf {\bibinfo {volume} {72}},\ \bibinfo {pages} {2400036} (\bibinfo {year} {2024}{\natexlab{b}})},\ \Eprint {http://arxiv.org/abs/2311.11650} {arXiv:2311.11650 [gr-qc]} \BibitemShut {NoStop}%
\bibitem [{\citenamefont {Skvortsova}(2024{\natexlab{c}})}]{Skvortsova:2024atk}%
  \BibitemOpen
  \bibfield  {author} {\bibinfo {author} {\bibfnamefont {M.}~\bibnamefont {Skvortsova}},\ }\href {\doibase 10.1002/prop.202400132} {\bibfield  {journal} {\bibinfo  {journal} {Fortsch. Phys.}\ }\textbf {\bibinfo {volume} {72}},\ \bibinfo {pages} {2400132} (\bibinfo {year} {2024}{\natexlab{c}})},\ \Eprint {http://arxiv.org/abs/2405.06390} {arXiv:2405.06390 [gr-qc]} \BibitemShut {NoStop}%
\bibitem [{\citenamefont {Zinhailo}(2018)}]{Zinhailo:2018ska}%
  \BibitemOpen
  \bibfield  {author} {\bibinfo {author} {\bibfnamefont {A.~F.}\ \bibnamefont {Zinhailo}},\ }\href {\doibase 10.1140/epjc/s10052-018-6467-8} {\bibfield  {journal} {\bibinfo  {journal} {Eur. Phys. J. C}\ }\textbf {\bibinfo {volume} {78}},\ \bibinfo {pages} {992} (\bibinfo {year} {2018})},\ \Eprint {http://arxiv.org/abs/1809.03913} {arXiv:1809.03913 [gr-qc]} \BibitemShut {NoStop}%
\bibitem [{\citenamefont {Churilova}\ \emph {et~al.}(2021)\citenamefont {Churilova}, \citenamefont {Konoplya}, \citenamefont {Stuchlik},\ and\ \citenamefont {Zhidenko}}]{Churilova:2021tgn}%
  \BibitemOpen
  \bibfield  {author} {\bibinfo {author} {\bibfnamefont {M.~S.}\ \bibnamefont {Churilova}}, \bibinfo {author} {\bibfnamefont {R.~A.}\ \bibnamefont {Konoplya}}, \bibinfo {author} {\bibfnamefont {Z.}~\bibnamefont {Stuchlik}}, \ and\ \bibinfo {author} {\bibfnamefont {A.}~\bibnamefont {Zhidenko}},\ }\href {\doibase 10.1088/1475-7516/2021/10/010} {\bibfield  {journal} {\bibinfo  {journal} {JCAP}\ }\textbf {\bibinfo {volume} {10}},\ \bibinfo {pages} {010} (\bibinfo {year} {2021})},\ \Eprint {http://arxiv.org/abs/2107.05977} {arXiv:2107.05977 [gr-qc]} \BibitemShut {NoStop}%
\bibitem [{\citenamefont {Bronnikov}\ \emph {et~al.}(2021)\citenamefont {Bronnikov}, \citenamefont {Konoplya},\ and\ \citenamefont {Pappas}}]{Bronnikov:2021liv}%
  \BibitemOpen
  \bibfield  {author} {\bibinfo {author} {\bibfnamefont {K.~A.}\ \bibnamefont {Bronnikov}}, \bibinfo {author} {\bibfnamefont {R.~A.}\ \bibnamefont {Konoplya}}, \ and\ \bibinfo {author} {\bibfnamefont {T.~D.}\ \bibnamefont {Pappas}},\ }\href {\doibase 10.1103/PhysRevD.103.124062} {\bibfield  {journal} {\bibinfo  {journal} {Phys. Rev. D}\ }\textbf {\bibinfo {volume} {103}},\ \bibinfo {pages} {124062} (\bibinfo {year} {2021})},\ \Eprint {http://arxiv.org/abs/2102.10679} {arXiv:2102.10679 [gr-qc]} \BibitemShut {NoStop}%
\bibitem [{\citenamefont {Han}\ and\ \citenamefont {Gwak}(2025)}]{Han:2025cal}%
  \BibitemOpen
  \bibfield  {author} {\bibinfo {author} {\bibfnamefont {H.}~\bibnamefont {Han}}\ and\ \bibinfo {author} {\bibfnamefont {B.}~\bibnamefont {Gwak}},\ }\href@noop {} {\  (\bibinfo {year} {2025})},\ \Eprint {http://arxiv.org/abs/2508.12989} {arXiv:2508.12989 [gr-qc]} \BibitemShut {NoStop}%
\bibitem [{\citenamefont {Bolokhov}\ and\ \citenamefont {Skvortsova}(2025{\natexlab{b}})}]{Bolokhov:2024otn}%
  \BibitemOpen
  \bibfield  {author} {\bibinfo {author} {\bibfnamefont {S.~V.}\ \bibnamefont {Bolokhov}}\ and\ \bibinfo {author} {\bibfnamefont {M.}~\bibnamefont {Skvortsova}},\ }\href {\doibase 10.1088/1475-7516/2025/04/025} {\bibfield  {journal} {\bibinfo  {journal} {JCAP}\ }\textbf {\bibinfo {volume} {04}},\ \bibinfo {pages} {025} (\bibinfo {year} {2025}{\natexlab{b}})},\ \Eprint {http://arxiv.org/abs/2412.11166} {arXiv:2412.11166 [gr-qc]} \BibitemShut {NoStop}%
\bibitem [{\citenamefont {Skvortsova}(2025{\natexlab{b}})}]{Skvortsova:2024msa}%
  \BibitemOpen
  \bibfield  {author} {\bibinfo {author} {\bibfnamefont {M.}~\bibnamefont {Skvortsova}},\ }\href {\doibase 10.1140/epjc/s10052-025-14589-w} {\bibfield  {journal} {\bibinfo  {journal} {Eur. Phys. J. C}\ }\textbf {\bibinfo {volume} {85}},\ \bibinfo {pages} {854} (\bibinfo {year} {2025}{\natexlab{b}})},\ \Eprint {http://arxiv.org/abs/2411.06007} {arXiv:2411.06007 [gr-qc]} \BibitemShut {NoStop}%
\bibitem [{\citenamefont {Malik}(2025{\natexlab{c}})}]{Malik:2024cgb}%
  \BibitemOpen
  \bibfield  {author} {\bibinfo {author} {\bibfnamefont {Z.}~\bibnamefont {Malik}},\ }\href {\doibase 10.1088/1475-7516/2025/04/042} {\bibfield  {journal} {\bibinfo  {journal} {JCAP}\ }\textbf {\bibinfo {volume} {04}},\ \bibinfo {pages} {042} (\bibinfo {year} {2025}{\natexlab{c}})},\ \Eprint {http://arxiv.org/abs/2412.19443} {arXiv:2412.19443 [gr-qc]} \BibitemShut {NoStop}%
\bibitem [{\citenamefont {L{\"u}tf{\"u}o{\u{g}}lu}(2025{\natexlab{b}})}]{Lutfuoglu:2025hjy}%
  \BibitemOpen
  \bibfield  {author} {\bibinfo {author} {\bibfnamefont {B.~C.}\ \bibnamefont {L{\"u}tf{\"u}o{\u{g}}lu}},\ }\href {\doibase 10.1140/epjc/s10052-025-14210-0} {\bibfield  {journal} {\bibinfo  {journal} {Eur. Phys. J. C}\ }\textbf {\bibinfo {volume} {85}},\ \bibinfo {pages} {486} (\bibinfo {year} {2025}{\natexlab{b}})},\ \Eprint {http://arxiv.org/abs/2503.16087} {arXiv:2503.16087 [gr-qc]} \BibitemShut {NoStop}%
\bibitem [{\citenamefont {L{\"u}tf{\"u}o{\u{g}}lu}\ \emph {et~al.}(2025)\citenamefont {L{\"u}tf{\"u}o{\u{g}}lu}, \citenamefont {Saka}, \citenamefont {Shermatov}, \citenamefont {Rayimbaev}, \citenamefont {Ibragimov},\ and\ \citenamefont {Muminov}}]{Lutfuoglu:2025blw}%
  \BibitemOpen
  \bibfield  {author} {\bibinfo {author} {\bibfnamefont {B.~C.}\ \bibnamefont {L{\"u}tf{\"u}o{\u{g}}lu}}, \bibinfo {author} {\bibfnamefont {E.~U.}\ \bibnamefont {Saka}}, \bibinfo {author} {\bibfnamefont {A.}~\bibnamefont {Shermatov}}, \bibinfo {author} {\bibfnamefont {J.}~\bibnamefont {Rayimbaev}}, \bibinfo {author} {\bibfnamefont {I.}~\bibnamefont {Ibragimov}}, \ and\ \bibinfo {author} {\bibfnamefont {S.}~\bibnamefont {Muminov}},\ }\href {\doibase 10.1140/epjc/s10052-025-14950-z} {\bibfield  {journal} {\bibinfo  {journal} {Eur. Phys. J. C}\ }\textbf {\bibinfo {volume} {85}},\ \bibinfo {pages} {1190} (\bibinfo {year} {2025})},\ \Eprint {http://arxiv.org/abs/2509.15923} {arXiv:2509.15923 [gr-qc]} \BibitemShut {NoStop}%
\bibitem [{\citenamefont {Malik}(2025{\natexlab{d}})}]{Malik:2025dxn}%
  \BibitemOpen
  \bibfield  {author} {\bibinfo {author} {\bibfnamefont {Z.}~\bibnamefont {Malik}},\ }\href {\doibase 10.1007/s10773-025-06198-w} {\bibfield  {journal} {\bibinfo  {journal} {Int. J. Theor. Phys.}\ }\textbf {\bibinfo {volume} {64}},\ \bibinfo {pages} {314} (\bibinfo {year} {2025}{\natexlab{d}})},\ \Eprint {http://arxiv.org/abs/2508.19178} {arXiv:2508.19178 [gr-qc]} \BibitemShut {NoStop}%
\bibitem [{\citenamefont {Dubinsky}(2025)}]{Dubinsky:2024vbn}%
  \BibitemOpen
  \bibfield  {author} {\bibinfo {author} {\bibfnamefont {A.}~\bibnamefont {Dubinsky}},\ }\href {\doibase 10.1142/S0217732325501111} {\bibfield  {journal} {\bibinfo  {journal} {Mod. Phys. Lett. A}\ }\textbf {\bibinfo {volume} {40}},\ \bibinfo {pages} {2550111} (\bibinfo {year} {2025})},\ \Eprint {http://arxiv.org/abs/2412.00625} {arXiv:2412.00625 [gr-qc]} \BibitemShut {NoStop}%
\bibitem [{\citenamefont {Bolokhov}\ and\ \citenamefont {Skvortsova}(2025{\natexlab{c}})}]{Bolokhov:2025lnt}%
  \BibitemOpen
  \bibfield  {author} {\bibinfo {author} {\bibfnamefont {S.~V.}\ \bibnamefont {Bolokhov}}\ and\ \bibinfo {author} {\bibfnamefont {M.}~\bibnamefont {Skvortsova}},\ }\href@noop {} {\bibfield  {journal} {\bibinfo  {journal} {Int. J. Grav. Theor. Phys.}\ }\textbf {\bibinfo {volume} {1}},\ \bibinfo {pages} {3} (\bibinfo {year} {2025}{\natexlab{c}})},\ \Eprint {http://arxiv.org/abs/2507.07196} {arXiv:2507.07196 [gr-qc]} \BibitemShut {NoStop}%
\bibitem [{\citenamefont {L{\"u}tf{\"u}o{\u{g}}lu}(2026)}]{Lutfuoglu:2025ohb}%
  \BibitemOpen
  \bibfield  {author} {\bibinfo {author} {\bibfnamefont {B.~C.}\ \bibnamefont {L{\"u}tf{\"u}o{\u{g}}lu}},\ }\href {\doibase 10.1140/epjc/s10052-026-15290-2} {\bibfield  {journal} {\bibinfo  {journal} {Eur. Phys. J. C}\ }\textbf {\bibinfo {volume} {86}},\ \bibinfo {pages} {39} (\bibinfo {year} {2026})},\ \Eprint {http://arxiv.org/abs/2505.06966} {arXiv:2505.06966 [gr-qc]} \BibitemShut {NoStop}%
\bibitem [{\citenamefont {L{\"u}tf{\"u}o{\u{g}}lu}(2025{\natexlab{c}})}]{Lutfuoglu:2025ldc}%
  \BibitemOpen
  \bibfield  {author} {\bibinfo {author} {\bibfnamefont {B.~C.}\ \bibnamefont {L{\"u}tf{\"u}o{\u{g}}lu}},\ }\href {\doibase 10.53941/ijgtp.2025.100004} {\bibfield  {journal} {\bibinfo  {journal} {Int. J. Grav. Theor. Phys.}\ }\textbf {\bibinfo {volume} {1}},\ \bibinfo {pages} {4} (\bibinfo {year} {2025}{\natexlab{c}})},\ \Eprint {http://arxiv.org/abs/2507.09246} {arXiv:2507.09246 [gr-qc]} \BibitemShut {NoStop}%
\bibitem [{\citenamefont {Dubinsky}(2026)}]{Dubinsky:2025nxv}%
  \BibitemOpen
  \bibfield  {author} {\bibinfo {author} {\bibfnamefont {A.}~\bibnamefont {Dubinsky}},\ }\href {\doibase 10.1016/j.aop.2025.170299} {\bibfield  {journal} {\bibinfo  {journal} {Annals Phys.}\ }\textbf {\bibinfo {volume} {485}},\ \bibinfo {pages} {170299} (\bibinfo {year} {2026})},\ \Eprint {http://arxiv.org/abs/2509.11017} {arXiv:2509.11017 [gr-qc]} \BibitemShut {NoStop}%
\bibitem [{\citenamefont {Malik}(2025{\natexlab{e}})}]{Malik:2025erb}%
  \BibitemOpen
  \bibfield  {author} {\bibinfo {author} {\bibfnamefont {Z.}~\bibnamefont {Malik}},\ }\href {\doibase 10.53941/ijgtp.2025.100006} {\bibfield  {journal} {\bibinfo  {journal} {Int. J. Grav. Theor. Phys.}\ }\textbf {\bibinfo {volume} {1}},\ \bibinfo {pages} {6} (\bibinfo {year} {2025}{\natexlab{e}})},\ \Eprint {http://arxiv.org/abs/2509.15995} {arXiv:2509.15995 [gr-qc]} \BibitemShut {NoStop}%
\bibitem [{\citenamefont {Konoplya}\ and\ \citenamefont {Zhidenko}(2017)}]{Konoplya:2017lhs}%
  \BibitemOpen
  \bibfield  {author} {\bibinfo {author} {\bibfnamefont {R.~A.}\ \bibnamefont {Konoplya}}\ and\ \bibinfo {author} {\bibfnamefont {A.}~\bibnamefont {Zhidenko}},\ }\href {\doibase 10.1088/1475-7516/2017/05/050} {\bibfield  {journal} {\bibinfo  {journal} {JCAP}\ }\textbf {\bibinfo {volume} {05}},\ \bibinfo {pages} {050} (\bibinfo {year} {2017})},\ \Eprint {http://arxiv.org/abs/1705.01656} {arXiv:1705.01656 [hep-th]} \BibitemShut {NoStop}%
\bibitem [{\citenamefont {Konoplya}\ and\ \citenamefont {Stuchlík}(2017)}]{Konoplya:2017wot}%
  \BibitemOpen
  \bibfield  {author} {\bibinfo {author} {\bibfnamefont {R.~A.}\ \bibnamefont {Konoplya}}\ and\ \bibinfo {author} {\bibfnamefont {Z.}~\bibnamefont {Stuchlík}},\ }\href {\doibase 10.1016/j.physletb.2017.06.015} {\bibfield  {journal} {\bibinfo  {journal} {Phys. Lett. B}\ }\textbf {\bibinfo {volume} {771}},\ \bibinfo {pages} {597} (\bibinfo {year} {2017})},\ \Eprint {http://arxiv.org/abs/1705.05928} {arXiv:1705.05928 [gr-qc]} \BibitemShut {NoStop}%
\bibitem [{\citenamefont {Konoplya}\ and\ \citenamefont {Zhidenko}(2004)}]{Konoplya:2004uk}%
  \BibitemOpen
  \bibfield  {author} {\bibinfo {author} {\bibfnamefont {R.~A.}\ \bibnamefont {Konoplya}}\ and\ \bibinfo {author} {\bibfnamefont {A.}~\bibnamefont {Zhidenko}},\ }\href {\doibase 10.1088/1126-6708/2004/06/037} {\bibfield  {journal} {\bibinfo  {journal} {JHEP}\ }\textbf {\bibinfo {volume} {06}},\ \bibinfo {pages} {037} (\bibinfo {year} {2004})},\ \Eprint {http://arxiv.org/abs/hep-th/0402080} {arXiv:hep-th/0402080} \BibitemShut {NoStop}%
\bibitem [{\citenamefont {Zhidenko}(2004)}]{Zhidenko:2003wq}%
  \BibitemOpen
  \bibfield  {author} {\bibinfo {author} {\bibfnamefont {A.}~\bibnamefont {Zhidenko}},\ }\href {\doibase 10.1088/0264-9381/21/1/019} {\bibfield  {journal} {\bibinfo  {journal} {Class. Quant. Grav.}\ }\textbf {\bibinfo {volume} {21}},\ \bibinfo {pages} {273} (\bibinfo {year} {2004})},\ \Eprint {http://arxiv.org/abs/gr-qc/0307012} {arXiv:gr-qc/0307012} \BibitemShut {NoStop}%
\bibitem [{\citenamefont {Moss}\ and\ \citenamefont {Norman}(2002)}]{Moss:2001ga}%
  \BibitemOpen
  \bibfield  {author} {\bibinfo {author} {\bibfnamefont {I.~G.}\ \bibnamefont {Moss}}\ and\ \bibinfo {author} {\bibfnamefont {J.~P.}\ \bibnamefont {Norman}},\ }\href {\doibase 10.1088/0264-9381/19/8/319} {\bibfield  {journal} {\bibinfo  {journal} {Class. Quant. Grav.}\ }\textbf {\bibinfo {volume} {19}},\ \bibinfo {pages} {2323} (\bibinfo {year} {2002})},\ \Eprint {http://arxiv.org/abs/gr-qc/0201016} {arXiv:gr-qc/0201016} \BibitemShut {NoStop}%
\bibitem [{\citenamefont {Lopez-Ortega}(2007)}]{Lopez-Ortega:2007vlo}%
  \BibitemOpen
  \bibfield  {author} {\bibinfo {author} {\bibfnamefont {A.}~\bibnamefont {Lopez-Ortega}},\ }\href {\doibase 10.1007/s10714-007-0435-1} {\bibfield  {journal} {\bibinfo  {journal} {Gen. Rel. Grav.}\ }\textbf {\bibinfo {volume} {39}},\ \bibinfo {pages} {1011} (\bibinfo {year} {2007})},\ \Eprint {http://arxiv.org/abs/0704.2468} {arXiv:0704.2468 [gr-qc]} \BibitemShut {NoStop}%
\bibitem [{\citenamefont {Lopez-Ortega}(2012)}]{Lopez-Ortega:2012xvr}%
  \BibitemOpen
  \bibfield  {author} {\bibinfo {author} {\bibfnamefont {A.}~\bibnamefont {Lopez-Ortega}},\ }\href {\doibase 10.1007/s10714-012-1398-4} {\bibfield  {journal} {\bibinfo  {journal} {Gen. Rel. Grav.}\ }\textbf {\bibinfo {volume} {44}},\ \bibinfo {pages} {2387} (\bibinfo {year} {2012})},\ \Eprint {http://arxiv.org/abs/1207.6791} {arXiv:1207.6791 [gr-qc]} \BibitemShut {NoStop}%
\bibitem [{\citenamefont {Konoplya}(2023{\natexlab{b}})}]{Konoplya:2022gjp}%
  \BibitemOpen
  \bibfield  {author} {\bibinfo {author} {\bibfnamefont {R.~A.}\ \bibnamefont {Konoplya}},\ }\href {\doibase 10.1016/j.physletb.2023.137674} {\bibfield  {journal} {\bibinfo  {journal} {Phys. Lett. B}\ }\textbf {\bibinfo {volume} {838}},\ \bibinfo {pages} {137674} (\bibinfo {year} {2023}{\natexlab{b}})},\ \Eprint {http://arxiv.org/abs/2210.08373} {arXiv:2210.08373 [gr-qc]} \BibitemShut {NoStop}%
\bibitem [{\citenamefont {Konoplya}\ \emph {et~al.}(2007)\citenamefont {Konoplya}, \citenamefont {Zhidenko},\ and\ \citenamefont {Molina}}]{Konoplya:2006gq}%
  \BibitemOpen
  \bibfield  {author} {\bibinfo {author} {\bibfnamefont {R.~A.}\ \bibnamefont {Konoplya}}, \bibinfo {author} {\bibfnamefont {A.}~\bibnamefont {Zhidenko}}, \ and\ \bibinfo {author} {\bibfnamefont {C.}~\bibnamefont {Molina}},\ }\href {\doibase 10.1103/PhysRevD.75.084004} {\bibfield  {journal} {\bibinfo  {journal} {Phys. Rev. D}\ }\textbf {\bibinfo {volume} {75}},\ \bibinfo {pages} {084004} (\bibinfo {year} {2007})},\ \Eprint {http://arxiv.org/abs/gr-qc/0602047} {arXiv:gr-qc/0602047} \BibitemShut {NoStop}%
\bibitem [{\citenamefont {Ohashi}\ and\ \citenamefont {Sakagami}(2004)}]{Ohashi:2004wr}%
  \BibitemOpen
  \bibfield  {author} {\bibinfo {author} {\bibfnamefont {A.}~\bibnamefont {Ohashi}}\ and\ \bibinfo {author} {\bibfnamefont {M.-a.}\ \bibnamefont {Sakagami}},\ }\href {\doibase 10.1088/0264-9381/21/16/010} {\bibfield  {journal} {\bibinfo  {journal} {Class. Quant. Grav.}\ }\textbf {\bibinfo {volume} {21}},\ \bibinfo {pages} {3973} (\bibinfo {year} {2004})},\ \Eprint {http://arxiv.org/abs/gr-qc/0407009} {arXiv:gr-qc/0407009} \BibitemShut {NoStop}%
\bibitem [{\citenamefont {Futterman}\ \emph {et~al.}(2012)\citenamefont {Futterman}, \citenamefont {Handler},\ and\ \citenamefont {Matzner}}]{Futterman:1988ni}%
  \BibitemOpen
  \bibfield  {author} {\bibinfo {author} {\bibfnamefont {J.~A.~H.}\ \bibnamefont {Futterman}}, \bibinfo {author} {\bibfnamefont {F.~A.}\ \bibnamefont {Handler}}, \ and\ \bibinfo {author} {\bibfnamefont {R.~A.}\ \bibnamefont {Matzner}},\ }\href {\doibase 10.1017/CBO9780511735615} {\emph {\bibinfo {title} {{SCATTERING FROM BLACK HOLES}}}},\ Cambridge Monographs on Mathematical Physics\ (\bibinfo  {publisher} {Cambridge University Press},\ \bibinfo {year} {2012})\BibitemShut {NoStop}%
\end{thebibliography}%
\end{document}